\documentclass{article}
\usepackage{graphicx} 
\usepackage{fullpage}
\usepackage{diagbox}
\usepackage{amsmath, amssymb}
\usepackage{cancel}
\usepackage{bbm}
\usepackage{tikz}
\usepackage{algorithm}
\usepackage{algorithmic}
\usepackage{url}
\usepackage[colorlinks=true, allcolors=blue]{hyperref}

\newcommand{\eps}{\epsilon}
\DeclareMathOperator{\EX}{\mathbb{E}}
\newtheorem{theorem}{Theorem}[section]

\newtheorem{corollary}[theorem]{Corollary}
\newenvironment{theorem*}[1][]{%
  \trivlist\item[\hskip\labelsep{\bfseries Theorem\ (#1).}]\itshape
}{\endtrivlist}
\newtheorem{lemma}[theorem]{Lemma}
\newtheorem{claim}[theorem]{Claim}

\newtheorem{example}{Example}[section]
\newtheorem{observation}[theorem]{Observation}
\newenvironment{proof}[1][Proof]{\noindent {\bf #1:} \hspace{.677em}}
                      {\hspace*{\fill}{\eod}}

\def\eod{\vrule height 6pt width 5pt depth 0pt}

\usepackage[]{color-edits}
\addauthor{TODO}{red}

\addauthor{ITC}{magenta}
\addauthor{SR}{blue}
\addauthor{VG}{red}

\title{Sequential Contract Design for Budgeted Crowdsourcing}
\author{
Michal Feldman$^{1}$ \qquad Vasilis Gkatzelis$^{2}$\\
Shaul Rosner$^{1}$ \qquad Inbal Talgam-Cohen$^{1}$\\[0.8em]
{\small $^{1}$Blavatnik School of Computer Science and AI, Tel Aviv University, Tel Aviv, Israel}\\
{\small $^{2}$College of Computing and Informatics, Drexel University, Philadelphia, USA}
}
\date{}

\begin{document}

\maketitle

\begin{abstract}
    A principal wishes to crowdsource a collection of tasks under a fixed compensation budget. The quality of a task’s outcome is stochastic and depends on the worker’s effort level. Since this effort is unobservable, the principal offers contracts to workers specifying their compensation for each possible outcome. The principal offers these contracts sequentially, using previously observed outcomes and payments to decide whom to approach next and what contract to offer. The objective is to maximize the expected value of the work produced while ensuring that total payments never exceed the budget, regardless of the realized outcomes. Contract design must therefore account both for a worker’s incentives and for the budget available to incentivize others.

For valuations that are additive across tasks, we characterize how the approximability of the optimal adaptive policy depends on the numbers of workers, effort levels, and outcomes. Polynomial-time constant-factor approximation is possible whenever the number of workers, effort levels per worker, or outcomes per task is constant, but impossible in general when all three are unrestricted, unless  $P=NP$. We obtain tight guarantees for every fixed number of workers, establish computational hardness even in restricted settings, and obtain exact algorithms and approximation schemes in more structured settings. 

We also investigate simple contract structures, demonstrating both the benefits of adapting contracts to previous outcomes and the limitations of local contract rules, while identifying conditions under which a local policy achieves a constant-factor guarantee. Finally, we extend our analysis beyond additive valuations, obtaining constant-factor approximations for monotone submodular valuations with binary outcomes, including guarantees against a stronger full-knowledge benchmark under additional assumptions.

\end{abstract}

    \thispagestyle{empty}
    \pagestyle{empty}
    \newpage
    \setcounter{tocdepth}{2} 
    \tableofcontents 
    \newpage
    \pagestyle{plain}
    \setcounter{page}{1}

\section{Introduction}
\label{sec:intro}

The same AI systems that have automated much of traditional crowdwork now depend on a steady supply of human input that cannot be automated. A model can summarize what has already been written, but it cannot report what happened at a city council meeting last night. It cannot confirm that a clinic listed online is still open, or judge whether a translation into a regional dialect sounds natural to native speakers. As AI-generated text saturates the web, freshly created and independently validated human content is becoming a scarce input. Organizations that build or deploy AI systems increasingly need to budget for it directly, by paying contributors to produce and check new data.

A central obstacle is that the quality of a contribution depends on how much effort the worker invests, and effort is costly. A careful fact-check requires tracing claims to primary sources. A useful field report requires actually visiting the site. A good expert annotation requires thought rather than a quick guess. Workers invest such effort only if they are compensated for it, and the requester’s ability to compensate is limited by a fixed budget. Every dollar committed to one worker is a dollar unavailable to incentivize another. The requester must therefore decide not only how much quality to pay for from each worker, but also which workers to pay at all.

Effort is typically unobservable, so the requester cannot pay for it directly. She can, however, pay according to the observed quality of the result, for instance through a bonus for higher quality outcomes. This is the classic principal-agent problem \cite{Holm79}: the principal commits to a contract specifying a payment for each possible outcome, and the agent chooses the effort level that maximizes their expected payment minus their cost. Most work on multi-agent contracts considers agents whose efforts jointly determine a single shared outcome \cite{BFNW12,DEFK25-mulmul}. In crowdsourcing, by contrast, each worker completes a separate task, and each produces an individual outcome that can be observed and compensated on its own \cite{CMG23}.

Paying for outcomes rather than effort has a consequence that is easy to overlook: the principal’s spending becomes a random variable. Suppose she knows exactly how each worker’s effort shapes the distribution over outcome quality. She still cannot know in advance how much she will end up paying, because a bonus for excellent work is paid only if excellent work materializes. A budget, however, is a hard constraint. Payments must remain within it for every realization, not just in expectation.

If effort were observable, the problem would reduce to knapsack: pay each chosen worker their cost and select the most valuable set that fits. With hidden effort, the payment a worker receives depends on the outcome their effort produces. Covering it ex post requires reserving budget for payments that may never be made. Even when values are additive across tasks, the budget creates a strong interaction between workers. How a contract spreads payments over outcomes affects not only one worker’s incentives but also how much budget is likely to remain for everyone else.

Sequential interaction offers a way around this. A principal who approaches workers one at a time sees each outcome and the resulting payment before deciding what to do next. We therefore study sequential contract design, where the principal repeatedly chooses which worker to approach and which contract to offer, based on the amount of budget that is still available. Our goal is to design sequential contracts that maximize the principal's expected value, and we use the optimal adaptive policy as our benchmark. Adaptivity matters along both dimensions: adapting the order in which agents are approached and adapting the contracts offered to them.
If we fix every contract in advance, the problem resembles stochastic knapsack \cite{DGV08}, with payments playing the role of random item sizes, but there are key differences: in our setting the principal chooses the size distribution through the contract, subject to incentive compatibility, and she must never overflow the budget.
\subsection{Our Contributions}
\label{sec:contributions}
We study the computational problem of maximizing the principal's value under an ex-post budget constraint. First, we consider how well the optimal adaptive policy can be approximated in the additive setting, depending on the number of agents $n$, the number of actions available to each agent (the effort levels), and the number of outcomes. Our approximation and hardness guarantees are summarized in Figure~\ref{fig:parameter-summary}. Next, we ask under which structural assumptions substantially simpler contracts can still provide good guarantees. Finally, we study the extension of our model beyond additive valuations.

We begin by considering settings with an arbitrary number of outcomes and a constant number of actions. The main difficulty is that the principal must jointly determine which agents to approach, which contracts to offer them, and how these choices depend on the remaining budget. 
We give a linear-programming relaxation that captures both the contract offered to an agent  and the amount of budget that may already have been spent when the agent is approached. The relaxation uses a discretization of possible budget thresholds and a restricted family of contracts, and is combined with randomized rounding to obtain our main approximation result:

\vspace{0.1in}
\noindent\textbf{Theorem}
(see Theorem~\ref{thm:lp_rounding}).
For every fixed $\epsilon>0$, there is a polynomial-time
$\frac{1}{16(1+\epsilon)}$-approximation to the optimal adaptive policy when the number of actions is constant.
Moreover, when the number of outcomes is constant, there is a polynomial-time
$\frac{1}{16}$-approximation for arbitrarily many actions.
\vspace{0.1in}

We remark that our approximation algorithm fixes the contracts and their order in advance, adapting only by skipping agents based on the remaining budget. This allows agents to be approached in parallel whenever their contracts are guaranteed to be offered regardless of the outcomes of the other agents in the same batch.

Notably, exact optimization
remains NP-hard even under severe restrictions on the
number of agents, actions, or outcomes.

\vspace{0.1in}
\noindent\textbf{Theorem}
(see Theorems~\ref{thm:2agent_hard} and~\ref{thm:binary_hard}).
Computing an optimal adaptive policy is NP-hard even with
two actions per agent, both when the number of outcomes
is two and the number of agents is arbitrary, and when
the number of agents is two and the number of outcomes
is arbitrary.
\vspace{0.1in}

We next consider limiting the number of agents instead of the number of actions or outcomes. When both the number of agents and the number of outcomes are constant, the optimal adaptive policy can be computed exactly in polynomial time (see Observation~\ref{obs:const-n-m}). When the number of agents and actions is constant but the number of outcomes is arbitrary, a polynomial-time $(1-\epsilon)$-approximation exists for every $\epsilon>0$ (see Theorem~\ref{thm:ptas}). Finally, when both the number of actions and the number of outcomes are arbitrary, a simple $\frac{1}{n}$-approximation is the best possible.

\vspace{0.1in}
\noindent\textbf{Theorem}
(see Theorems~\ref{thm:1agent_approx} and~\ref{thm:fixed_n_hardness}).
For any number $n$ of agents, there is a polynomial-time $\frac{1}{n}$-approximation to the optimal adaptive policy. Furthermore, for every fixed $n\geq2$ and every $\alpha\in(\frac{1}{n},1]$, for sufficiently many actions and outcomes, no polynomial-time $\alpha$-approximation exists unless $P=NP$, even for identical agents.
\vspace{0.1in}

The fixed-$n$ hardness result also implies that no constant-factor approximation is possible for an unrestricted number of agents when the number of actions and outcomes is also unrestricted. 


\vspace{0.1in}
\noindent\textbf{Theorem}
(see Corollary~\ref{cor:general_hardness}).
When both the number of actions and the number of outcomes are unrestricted, no polynomial-time constant-factor approximation to the optimal adaptive policy exists unless $P=NP$, even when the agents are identical.


\begin{figure}[h]
\centering
\scriptsize

\begin{minipage}{0.46\linewidth}
\centering
\textbf{Constant number of agents $n$}

\vspace{0.05in}
\begin{tabular}{c||c|c}
\hline
\diagbox[width=8em,height=3.5em]
  {\# actions}{\# outcomes}
    & $O(1)$ & arbitrary \\
\hline
\hline
$O(1)$
    & $\uparrow$ & $1-\eps$ \\
\hline
arbitrary
    & OPT & $1/n$ \\
\hline
\end{tabular}
\end{minipage}
\hfill
\begin{minipage}{0.46\linewidth}
\centering
\textbf{Arbitrary number of agents}

\vspace{0.05in}
\begin{tabular}{c||c|c}
\hline
\diagbox[width=8em,height=3.5em]
  {\# actions}{\# outcomes}
    & $O(1)$ & arbitrary \\
\hline
\hline
$O(1)$
    & $\uparrow$ & $\frac{1}{16(1+\epsilon)}$ \\
\hline
arbitrary
    & $\frac{1}{16}$ & no const. \\
\hline
\end{tabular}
\end{minipage}

\caption{Approximation guarantees as a function of the number of agents,
actions, and outcomes. Arrows indicate that the guarantee is inherited
from the cell pointed from.}
\label{fig:parameter-summary}
\end{figure}

\subsubsection*{Simple Contracts}

Our general algorithm chooses an agent's contract based on the full instance. We next ask whether substantially simpler contract structures can still provide good guarantees.

We first show that contract adaptivity itself has value. We construct an instance for which
$$
OPT_{\mathrm{static}}\leq \frac{5}{6}OPT,
$$
showing that adapting an agent's contract to previous outcomes can strictly improve the principal's value.

We next show that the differences between agents are not the only reason for the complexity of contracts. Even for identical agents, several natural simple contracts can perform arbitrarily poorly. In particular, any deterministic rule that determines an agent's contract solely from that agent's own parameters must assign the same contract to all identical agents, and we show that such contracts can be arbitrarily bad, which also rules out the minimax contract.

We identify a setting we term \emph{risk-free} agents, which are agents with binary action, which are guaranteed to deterministically achieve an outcome of value $0$ when shirking. With risk-free agents, considerably stronger simplicity guarantees are possible. For risk-free binary-action agents, we construct a local policy that assigns each agent its minimax contract and show that it provides a constant-factor approximation to the optimal adaptive policy.

\vspace{0.1in}
\noindent\textbf{Theorem} (see Theorem~\ref{thm:riskfree_gen}).
For risk-free binary-action agents, we can compute a policy using local contracts whose expected value is at least $\frac{1}{6}OPT$.
\vspace{0.1in}

We next consider another natural simple contract, the min-pay contract (Section \ref{sec:minpay}). For identical agents we prove it is optimal for two outcomes, and gives a constant-factor approximation for three outcomes, but can be arbitrarily bad for four or more outcomes. However, the failure of simple contracts in this setting is not due to computational hardness.

\vspace{0.1in}
\noindent\textbf{Theorem} (See Theorem~\ref{thm:identical_fptas}).
For every $\epsilon>0$, there is a polynomial-time $(1-\epsilon)$-approximation to the optimal adaptive policy for identical binary-action agents.
\vspace{0.1in}

\subsubsection*{Beyond Additive Valuations}

Finally, we move beyond additive valuations. To isolate the effect of non-additivity from the other difficulties of the model, we first consider risk-free binary-action agents with binary outcomes. For the basic AND and OR objectives, the optimal policy can still be computed efficiently (See Section \ref{sec:andor}). In passing, we also characterize the optimal policy for the classical
principal-utility objective in these settings.

We then consider monotone submodular valuations. Using a direct approach based on submodular maximization under a budget constraint, we obtain a constant-factor approximation even against the stronger full-knowledge benchmark.

\vspace{0.1in}
\noindent\textbf{Theorem} (See Theorem~\ref{thm:greedy_submod}).
For risk-free binary-action agents with binary outcomes and a monotone submodular valuation, there is a polynomial-time adaptive policy achieving a
$\frac{1}{4}(1-\frac{1}{e})$-approximation to the full-knowledge benchmark. If the required success payment is identical for all agents, the approximation improves to $1-\frac{1}{e}$.
\vspace{0.1in}

Finally, we show that the positive result for submodular valuations (against the standard benchmark) is not limited to the risk-free binary-action setting. Using stochastic probing, we obtain a constant-factor approximation after removing both assumptions.

\vspace{0.1in}
\noindent\textbf{Theorem} (See Theorem~\ref{thm:stochastic_probing}).
For binary-outcome agents with an arbitrary number of actions and a monotone submodular valuation, there is a polynomial-time constant-factor approximation to $OPT$.
\vspace{0.1in}
\subsection{Related Work}
\label{sec:related}
\emph{Contract Design.} Contract theory studies settings in which a principal delegates a task to an agent whose action is not directly observable. In the classical principal-agent model of Holmstr\"om \cite{Holm79} and Grossman and Hart \cite{GH83}, the principal commits to a contract specifying a payment for every possible outcome, and the agent chooses an action maximizing their expected payment minus the cost of the action. 
A natural extension considers a principal who relies on a \emph{team} of agents \cite{Holmstrom1982}, the hidden efforts of which jointly determine the outcome of a project. 

\emph{Algorithmic Contract Design.} 
In the combinatorial agency model of \cite{BFNW12}, the hidden efforts of the agents form substitutes and complements when determining the outcome. More recent work on multi-agent contracts studies richer value functions over the agents' outcomes, as well as settings in which every agent may choose among multiple actions \cite{DEFK26,DEFK25-mulmul}. Particularly relevant to our work is \cite{CMG23}, in which each agent's action independently determines an individual observable outcome. In~\cite{DoronSST25}, the agents have known sizes and must fit within a budget constraint. In these models, the principal faces not only the incentive problem of inducing desirable actions, but also a combinatorial problem of determining which agents to incentivize. 
This literature fits within a broader area of algorithmic contract design, see e.g.~\cite{DuttingRT19,DFTC24,Feldman25}.

\emph{Crowdsourcing and Incentives.}
Outcome-contingent contracts have also been studied specifically in crowdsourcing. \cite{HSV16,CohenDK22} study repeated crowdsourcing markets, where a requester adaptively adjusts quality-contingent payments and uses bandit algorithms to learn effective contracts. \cite{FLVZ23} design contracts for crowdsourced ranking, incentivizing accurate pairwise comparisons through the costly verification of a subset of the agents' responses. Unlike these works, we study the problem of allocating a shared compensation budget across multiple workers, assuming their costs and outcome distributions are known. Our focus is on adapting contracts to realized outcomes and the remaining budget, rather than on learning the workers' responses or verifying comparisons.

\emph{Sequential Contracting.}
Sequential interaction has appeared recently in several contract-design models, but in different forms. In \cite{EFS26} and \cite{HSS24}, a single agent sequentially chooses among several actions or exploration opportunities, while the principal designs the contract governing this process. 

Another recent direction studies online multi-agent contracts, where agents arrive gradually over time. \cite{LST26} considers adversarially arriving agents in an online hiring problem, while \cite{DFGK26} studies an online version of multi-agent contracts in which agents arrive one by one in a Bayesian setting.
Our setting differs in where the sequential decision lies. In our setting, the set of agents and their parameters are known to the principal, and it is the principal who chooses which agent to approach next. Moreover, the outcome of each interaction is observed before the next decision is made.

\emph{Contracting Under Budget Constraints.}
Several recent works study contract design under hard budget constraints \cite{FGPS25,AHT25}. In particular, these works consider multi-agent contract design with a budget on total payments and, among other objectives, maximizing the principal's reward subject to this constraint. A related model with individual binary outcomes is studied in \cite{GH26}, where a fixed reward budget is allocated among successful agents. \cite{ABLMTVZ26} studies compensation design with a budget-constrained principal and private agent costs, rather than known costs and hidden actions. More broadly, hard budget constraints are central to budget-feasible mechanism design, beginning with \cite{Singer10}. Budget-feasible procurement has also been studied through dynamic clock auctions, which adapt the prices offered to agents over multiple rounds \cite{BGGST22}. Budget-feasible mechanism design was also recently extended to study the design of self-funded procurement mechanisms, where the available budget is determined by the subset of agents that are hired~\cite{deng2026forging}.

A distinguishing feature of our setting is that the budget is consumed sequentially by stochastic, outcome-dependent payments. Feasibility must hold ex-post, so every payment promised by a contract must be covered by the budget remaining after the realized history.

\emph{Stochastic Knapsack.}
Our model is closely related to stochastic knapsack \cite{DGV08}, where items have stochastic sizes revealed upon selection. In our setting, realized payments play the role of stochastic sizes, while the principal's value is determined by the agents' realized outcomes.

There are several important differences between the two models. First, in our model an agent's value and payment are both determined by the same realized outcome and may therefore be correlated; related stochastic-knapsack models with correlated sizes and rewards have been studied in \cite{GKMR11}. Second, our budget constraint is ex-post, so a contract may only be offered if every payment it promises is affordable from the current budget, whereas classical stochastic knapsack permits an attempted item to overflow the remaining capacity \cite{DGV08}. Perhaps most importantly, the distribution of an item's size is fixed in stochastic knapsack, while in our model the principal \emph{chooses} the distribution of payments through the contract, subject to incentive compatibility.

A related model is studied in \cite{BBCEPT26}, where hidden effort affects the stochastic processing time of a job. They also develop skipping-adaptive approximation algorithms. In our setting, the contract instead determines the stochastic payment that both incentivizes the agent and consumes the budget.

\emph{Submodular Valuations and Stochastic Probing.}
Our results for non-additive valuations are related to classical work on monotone
submodular maximization. Maximizing a monotone submodular function subject to a
knapsack constraint admits a $(1-\frac{1}{e})$-approximation \cite{Svir04}, and
this guarantee is tight \cite{Feige98}. Our direct algorithm for risk-free
binary-action agents also uses standard guarantees for greedy submodular
maximization \cite{NWF78,KMN99}.

For general binary-outcome agents, we instead use the stochastic probing
framework, introduced in \cite{GN13}. Our result applies the framework of
\cite{FSZ21}, which gives constant-factor approximations for monotone submodular
stochastic probing under broad classes of inner and outer feasibility constraints.

\section{Model and Preliminaries}
\label{sec:prelim}
There are $n$ agents, each with $\ell$ actions and $m$ possible individual outcomes $o_1,\ldots,o_m$. If agent $i$ chooses action $a\in[\ell]$, outcome $o_j$ is realized with probability $p^i_{a,j}$. Action $a$ has cost $c^i_a\geq 0$, and we assume that every agent has a zero-cost action. Outcomes are realized independently across agents, conditioned on their actions.

We focus on additive valuations, where the principal's value is additive across agents, and extend our analysis to non-additive valuations in Section~\ref{sec:beyond_additive}. We denote by $r^i_j\in\mathbb{R}$ the principal's value from outcome $o_j$ of agent $i$. Let $a_i^0$ be the zero-cost action selected under the zero contract. We work with normalized values, satisfying $\sum_{j=1}^m p^i_{a_i^0,j}r^i_j=0.$

Thus, the principal's value from an outcome profile $(o_{j_1},\ldots,o_{j_n})$ is $\sum_{i=1}^n r^i_{j_i}$. This normalization is without loss of generality. Indeed, given arbitrary nonnegative outcome values $\widehat r^i_j$, we can define $r^i_j=\widehat r^i_j-\sum_{k=1}^m p^i_{a_i^0,k}\widehat r^i_k$. The subtracted baseline is independent of the policy and is nonnegative, so every multiplicative approximation guarantee for the normalized value also applies to the original total value. Note that the normalized outcome values may be negative. For each agent $i$ and action $a$, let $v^i_a=\sum_j p^i_{a,j}r^i_j$ denote the principal's expected additional value from that action.

An agent is \emph{binary-action} if it has two actions. We refer to Action $1$, which is selected under the zero contract and has zero cost, as \emph{shirking}, and to Action $2$ as \emph{exerting effort}. For binary-action agents, we write $c_i:=c^i_2$ and $v_i:=v^i_2$.

Our positive results for local contracts, as well as our
initial results for non-additive valuations, rely on \emph{risk-free} agents. A binary-action agent is risk-free if shirking
deterministically results in outcome $o_1$, so $p^i_{1,1}=1$. This captures settings in which shirking cannot produce a successful outcome, as is natural for crowdsourcing tasks where a worker must exert effort to complete the task. Recall that in
the additive setting $o_1$ has value $0$ by our normalization.

Agents are \emph{identical} if their action costs, outcome distributions, and outcome values are identical. For identical agents, we omit the agent index $i$.

\emph{Contracts and Incentives.}
The principal observes the realized outcome of an agent, but not the action chosen by the agent. A contract for agent $i$ is a vector $t^i=(t^i_1,\ldots,t^i_m)\in\mathbb{R}^m_{\geq 0}$ where $t^i_j$ is the payment made if outcome $o_j$ is realized. Thus, we assume limited liability.

Under contract $t^i$, agent $i$'s utility from action $a$ is $\sum_j p^i_{a,j}t^i_j-c^i_a$. A contract $t^i$ \emph{incentivizes} action $a$ if, for every $a'\in[\ell]$, $\sum_j(p^i_{a,j}-p^i_{a',j})t^i_j\geq c^i_a-c^i_{a'}$.

As is standard, we assume ties are broken in favor of the principal. Since every agent has a zero-cost action and payments are nonnegative, individual rationality is implied by incentive compatibility.

\emph{The Principal's Problem.}
The principal has a total budget $B$. She approaches agents sequentially in some order, and offers each agent a contract.  After observing the outcome of an agent, the principal may choose both the next agent and the contract offered to that agent. We treat an agent that is not approached as receiving the zero contract.

The principal's budget constraint must hold ex post. In particular, if the remaining budget at some point is $b$, a contract $t^i$ for agent $i$ may only be offered if $
    t^i_j\leq b$ for every $j\in[m]$.
After outcome $o_j$ is realized, the remaining budget becomes $b-t^i_j$. The principal's objective is to maximize her expected value subject to this constraint. We denote the maximum expected value achievable by an adaptive policy by $OPT$.

\section{General Additive Valuations}
\label{sec:general}
In this section we consider general additive valuations. In Section \ref{sec:main_result} we provide our main result, which is a constant-factor approximation when either the number of actions per agent, or the number of outcomes, is constant. In Section \ref{sec:const} we provide algorithms for instances with a constant number of agents. In Section \ref{sec:hardness} we provide our hardness results.

\subsection{Our Main Result}
\label{sec:main_result}
We first establish constant-factor approximation guarantees for arbitrarily many agents when either the number of actions or the number of outcomes is constant. 

\begin{theorem}
\label{thm:lp_rounding}
For every fixed $\epsilon>0$, there is a polynomial-time
$\frac{1}{16(1+\epsilon)}$-approximation to the optimal adaptive policy when $\ell=O(1)$.
For $m=O(1)$, there is a polynomial-time
$\frac{1}{16}$-approximation for arbitrarily many actions.
\end{theorem}

For simplicity, assume $B=1$ without loss of generality. Recall that we defined
$v^i_a =\sum_jp^i_{a,j}r^i_j$ as the principal's additional
expected value. We ignore actions with $v^i_a\leq 0$, as well
as actions that cannot be incentivized even with the full budget.

For every agent $i$, action $a$ and $d \in [0,1)$, consider the polytope of contracts which
incentivize action $a$ and pay at most $1-d$ for every outcome, given by $0\leq t^i_j\leq1-d$ for every $j$ and the IC constraint $\sum_jp^i_{a,j}t^i_j-c^i_a
\geq \sum_jp^i_{a',j}t^i_j-c^i_{a'}$ for every $a'\neq a$.

If at most $d$ of the budget has been
spent, then the remaining budget is at least $1-d$, so a contract in this polytope is ex-post budget
feasible. We associate such a contract with \emph{deadline} $d$, meaning that it is offered only while
at most $d$ of the budget has been spent. We say that such a contract is \emph{rigid} if its
payments are uniquely determined by $m$ linearly
independent tight constraints of this polytope.
Since there are at most $\ell-1$ incentive constraints,
a rigid contract pays $0$ or $1-d$ on all but at most
$\ell-1$ outcomes. Since every contract in the polytope is a convex combination of rigid
contracts, our linear program will use only these.

The payments of a rigid contract with deadline $0$ are determined by $m$ of its constraints
holding with equality, so they are the solution of a linear system whose entries come from the
instance. We choose $\eta>0$ such that every positive payment in a rigid contract with deadline $0$ is at least
$2\eta$, and such that $\log\frac{1}{\eta}$ is polynomial in the input size. Let $D = \{0\} \cup \{2^k\eta : k \geq 0,\ 2^k\eta<1\}$ be the set of deadlines we consider, so that $|D|$ is
polynomial in the input size.

For every agent $i$, action $a$, deadline
$d\in D$ and rigid contract $t^i$
incentivizing $a$, we define the variable
$x^i_{a,d,t^i}$, representing the probability
of incentivizing agent $i$ to choose $a$
using this contract. Consider the following linear program.

\begin{align*}
    \max\quad & \sum_{i,a,d,t^i} v^i_a x^i_{a,d,t^i}\\
    \text{s.t.}\quad
    & \sum_{a,d,t^i}x^i_{a,d,t^i} \leq 1
    && \forall i\\
    & \sum_{i,a}\sum_{\substack{d'\in D\\d'\leq d}}
      \sum_{t^i}x^i_{a,d',t^i}\\
    & \qquad\cdot\sum_j p^i_{a,j}
      \min\{t^i_j,2d,1\}\leq 4d
    && \forall d\in D\setminus\{0\}\\
    & \sum_{i,a,t^i}x^i_{a,0,t^i}
      \sum_{j:t^i_j>0}p^i_{a,j}\leq 2\\
    & x^i_{a,d,t^i}\geq 0
    && \forall i,a,d,t^i
\end{align*}

The first constraint reflects that an agent is incentivized at most once. The second limits the
total payment made by the agents whose deadline is at most $d$, where every payment $t^i_j$ is
replaced by $\min\{t^i_j,2d,1\}$. As $D$ is geometric with ratio $2$, an agent whose
deadline is at most $d$ is incentivized while the used budget is less than $2d$, and hence less
than $\min\{2d,1\}$, so in every realization the payments of all of them but the last sum to
less than $\min\{2d,1\}$, and the payment of the last one is replaced by at most
$\min\{2d,1\}$. Their total is therefore at most $2\min\{2d,1\} \leq 4d$. The third constraint
plays the same role for the agents whose deadline is $0$, where instead of replacing the
payments we count the agents which make a payment at all.

We denote by $LP$ the optimum of this linear program. Recall that $OPT$ is the expected value of the optimal adaptive policy.

\begin{lemma}
\label{lem:lp_upper_bound}
    $LP \geq OPT$.
\end{lemma}
\begin{proof}
    We construct a feasible solution of the linear program whose objective value is $OPT$, starting from a solution in which all variables are $0$. Fix an optimal adaptive algorithm, and consider a
    point in its execution at which it incentivizes agent $i$ to choose action $a$ with a contract $t^i$ when the
    remaining budget is $b$. If $b>1-\eta$ let $d=0$, and otherwise let $d$ be the largest element of $D$ with
    $d \leq 1-b$, and note that $1-b \leq 2d$. As the algorithm is
    ex-post budget feasible, $t^i_j \leq b \leq 1-d$ for every $j$, so $t^i$ lies in the polytope of
    $i$ for action $a$ with deadline $d$. We write $t^i$ as a convex combination $\sum_k \lambda_k t^{i,k}$ of
    the rigid contracts with deadline $d$, and increase the value of each variable $x^i_{a,d,t^{i,k}}$ by $\lambda_k$
    times the probability that the algorithm reaches this point.

    Every $t^{i,k}$ incentivizes action $a$, so the objective value of the resulting solution is
    the expected value of the optimal algorithm. Summing over the points at which agent $i$ is
    incentivized, $\sum_{a,d,t^i}x^i_{a,d,t^i}$ is the probability that the algorithm incentivizes
    agent $i$, which satisfies the first constraint. For the remaining constraints, note that for
    every $z \geq 0$ the function $\sum_j p^i_{a,j}\min\{t^i_j,z\}$ is concave in $t^i$, so
    $$\sum_k \lambda_k \sum_j p^i_{a,j}\min\{t^{i,k}_j,z\} \leq \sum_j p^i_{a,j}
    \min\{t^i_j,z\}.$$
        It is therefore sufficient to verify the remaining constraints for the contracts offered by the fixed optimal adaptive algorithm.

Fix $d\in D\setminus\{0\}$. Every contract assigned a deadline $d'\leq d$ is offered while the spent budget is less than $\min\{2d,1\}$. Indeed, for $d'>0$ we have $1-b<2d'\leq 2d$, while for $d'=0$ we have $1-b<\eta\leq d$; in both cases, $1-b<1$.

Now fix a sequence of realizations, and let $i'$ be the last agent incentivized with such a contract. Immediately before incentivizing $i'$, the total budget spent is less than $\min\{2d,1\}$. Thus, the payments to all preceding agents with deadlines at most $d$ contribute less than this amount to the second LP constraint, while the payment to $i'$ contributes at most $\min\{2d,1\}$ due to truncation. The total is therefore at most $2\min\{2d,1\}\leq 4d$ on every realization (and zero if no such agent exists). Taking expectations establishes the constraint.

The third constraint depends only on whether a payment is positive, not on its amount. By the definition of $\eta$, every positive payment of a rigid contract with deadline $0$ is at least $2\eta$, so $\sum_j p^i_{a,j}\min\{t^i_j,\eta\}=\eta\sum_{j:t^i_j>0}p^i_{a,j}$. Thus, by the preceding concavity argument, it suffices to bound the expected sum of payments truncated at $\eta$ under the optimal adaptive algorithm. In every realization, the total payment preceding the last agent assigned deadline $0$ is less than $\eta$, while the last agent contributes at most $\eta$ after truncation. The truncated sum is therefore at most $2\eta$, and taking expectations and dividing by $\eta$ proves the third constraint.

\end{proof}

As the linear program has exponentially many variables, we solve it approximately by applying
the ellipsoid method to its dual. Let $y_i$ be the dual variable of the first constraint of
agent $i$, let $\lambda_d$ be that of the second constraint for $d \in D \setminus \{0\}$, and
let $\lambda_0$ be that of the third constraint. The dual is the following linear program.

\begin{align*}
    \min\quad & \sum_i y_i + \sum_{d\in D\setminus\{0\}}
    4d\lambda_d + 2\lambda_0\\
    \text{s.t.}\quad
    & y_i + \lambda_0\sum_{j:t^i_j>0}
    p^i_{a,j}\mathbf 1_{d=0}\\
    &\quad + \sum_{\substack{d'\in D\setminus\{0\}\\d'\geq d}}
    \lambda_{d'}\sum_j p^i_{a,j}
    \min\{t^i_j,2d',1\}\geq v^i_a
    && \forall i,a,d,t^i\\
    & y_i\geq 0 && \forall i\\
    & \lambda_d\geq 0 && \forall d\in D
\end{align*}

For a fixed agent $i$, action $a$ and deadline $d$, separating over these constraints amounts to finding a
rigid contract with deadline $d$ which incentivizes action $a$ minimizing $\sum_jp^i_{a,j}g(t^i_j)$, where
$$g(z) = \lambda_0\mathbf 1_{d=0,\, z>0} + \sum_{d' \in D \setminus \{0\},\, d' \geq d}
\lambda_{d'}\min\{z,2d',1\}.$$
As every $d \in D$ has polynomial bit complexity, the same argument as above gives some
$\eta'>0$ with $\log\frac{1}{\eta'}$ polynomial in the input size, such that every positive
payment in a rigid contract with a deadline in $D$ is at least $\eta'$.

For a candidate dual solution, the constraints of a fixed agent $i$, action $a$ and deadline $d$ differ only
in the rigid contract $t^i$ they refer to, so all of them hold if and only if the constraint of
the contract minimizing $\sum_jp^i_{a,j}g(t^i_j)$ holds. The following lemma finds such a
contract approximately, and we use it below to obtain a separation oracle for the dual.

\begin{lemma}
\label{lem:pricing}
For every agent $i$, action $a$ and deadline
$d\in D$, given non-negative dual variables,
a rigid contract incentivizing $a$ with
deadline $d$, minimizing
$\sum_jp^i_{a,j}g(t^i_j)$, can be computed:
\begin{enumerate}
    \item Within a factor of $1+\epsilon$ in
    polynomial time for $\ell=O(1)$ actions.
    \item Exactly in polynomial time for
    $m=O(1)$ outcomes.
\end{enumerate}
\end{lemma}
\begin{proof}
    As $g$ is concave, so is $\sum_jp^i_{a,j}g(t^i_j)$,
and therefore a minimizing rigid contract exists.
\paragraph{Constant number of actions.}For $\ell=1$, the zero contract is optimal, so assume $\ell\geq2$. Let $t$ be such a contract, and let $J$ be the set
of outcomes on which it pays more than
$0$ but less than $1-d$. By the definition of rigid contracts,
$|J|\leq\ell-1$.

    We enumerate $J$, and for each $j \in J$ choose a value $z_j$ from the geometric grid $\{\eta',(1+\epsilon)\eta',(1+\epsilon)^2\eta',\ldots\}$, restricted to values at most $1-d$, together with $1-d$ itself. For the correct guess,
$t_j\leq z_j\leq(1+\epsilon)t_j$.
By concavity and monotonicity of $g$,
$g(z_j)\leq(1+\epsilon)g(t_j)$.
As $|J|\leq\ell-1$, there are polynomially
many guesses for fixed $\ell$ and $\epsilon$.

        After guessing the outcomes with partial payments,
we need to determine which remaining outcomes receive
a payment of $1-d$. We will do so using a secondary linear program,
which may introduce additional partial payments.
To bound the cost of these additional payments,
we will first guess the outcomes with the largest
contributions to the pricing objective.

Formally, for guessing the outcomes with the largest contributions to the pricing objective, let $H$ contain the
$\lceil(\ell-1)/\epsilon\rceil$ outcomes with
the largest values of $p^i_{a,j}g(1-d)$ among
those paid $1-d$ by $t$, or all such outcomes
if there are fewer. We enumerate all possible
choices of $H$.

For every guess of $J$, the values $z_j$, and $H$, we solve the following secondary linear program:
\begin{align*}
\min\quad &
\sum_{j\notin J\cup H}p^i_{a,j}g(1-d)\frac{t_j}{1-d}\\
\text{s.t.}\quad &
\sum_j(p^i_{a,j}-p^i_{a',j})t_j
\geq c^i_a-c^i_{a'}
&& \forall a'\neq a,\\
& 0\leq t_j\leq z_j
&& \forall j\in J,\\
& t_j=1-d
&& \forall j\in H,\\
& 0\leq t_j\leq1-d
&& \forall j\notin J\cup H.
\end{align*}

By concavity and $g(0)=0$, the objective lower-bounds the pricing cost outside $J\cup H$, with equality for payments of $0$ or $1-d$. The constraints ensure incentive compatibility, fix the payments in $H$, and impose the guessed upper bounds on payments in $J$.

If $H$ contains the prescribed number of
outcomes, we additionally require $t_j=0$
for every $j\notin J\cup H$ whose value
$p^i_{a,j}g(1-d)$ exceeds the smallest
such value in $H$. Otherwise, we require
$t_j=0$ for every $j\notin J\cup H$.

Let $\hat t$ be an optimal solution determined by linearly independent tight constraints. For the correct guesses, $t$ is feasible for the secondary LP, so its linear objective at $\hat t$ is no greater than at $t$.
Since there are at most $\ell-1$ incentive
constraints, at most $\ell-1$ payments outside
$J$ are strictly between $0$ and $1-d$.

For outcomes in $J$, the cost is at most
$1+\epsilon$ times their cost in $t$.
The payments in $H$ are fixed, and outside $J\cup H$, the linear objective equals the actual cost whenever the payment is $0$ or $1-d$. The remaining payments can
increase the cost by at most $\ell-1$ times
the smallest value of $p^i_{a,j}g(1-d)$
in $H$. When $H$ contains the prescribed
number of outcomes, this additional cost
is at most $\epsilon\sum_{j\in H}p^i_{a,j}g(1-d)
\leq\epsilon\sum_jp^i_{a,j}g(t_j)$.

Otherwise, all payments outside $J\cup H$
are zero, so there is no additional cost.
Thus, $\hat t$ costs at most $1+2\epsilon$
times the optimum.

Finally, $\hat t$ has at most $2(\ell-1)$ payments strictly between $0$ and $1-d$. Fixing all other payments at $0$ or $1-d$, we can enumerate the feasible contracts of the original contract polytope directly using linearly independent tight constraints. Each such contract is rigid, and by concavity, at least one has a value of $\sum_j p^i_{a,j} g(t_j)$ no greater than its value at $\hat t$.

All enumerations take polynomial time
for fixed $\ell$ and $\epsilon$.
Since the construction achieves a factor of $1+2\epsilon$, applying it with parameter $\epsilon/2$ gives the claimed factor of $1+\epsilon$.

    \paragraph{Constant number of outcomes.}
When $m=O(1)$, the contract polytope has
$m$ variables and at most $2m+\ell-1$
constraints. Every rigid contract is determined
by $m$ linearly independent tight
constraints. We can therefore enumerate
at most $\binom{2m+\ell-1}{m}$ candidate systems, retain the feasible
rigid contracts, and return the one minimizing
$\sum_jp^i_{a,j}g(t^i_j)$.
For constant $m$, this takes polynomial
time, proving the claim.
\end{proof}

We use Lemma \ref{lem:pricing} with accuracy $\sqrt{1+\epsilon}-1$ to compute an approximate solution to the LP. Given
dual variables, we apply Lemma \ref{lem:pricing} to every agent $i$, action $a$ and deadline $d \in D$. If for some agent, action and deadline the contract it returns does not satisfy the corresponding dual
constraint, we return that constraint to the ellipsoid method. Otherwise, denoting by $C$ the minimum of $\sum_jp^i_{a,j}g(t^i_j)$ for agent $i$, action $a$, and
deadline $d$, the contract returned for them attains a value of at most $(\sqrt{1+\epsilon})C$, so
$y_i+\sqrt{1+\epsilon}C \geq v^i_a$ for every agent, action and deadline. Scaling the dual variables by $\sqrt{1+\epsilon}$ therefore gives a feasible solution of the dual.

We apply the ellipsoid method to the dual using the approximate separation oracle, with accuracy $\sqrt{1+\epsilon}-1$. Standard binary search on the dual objective bound yields a restricted primal LP, using only variables corresponding to dual constraints returned by the oracle, whose optimum $LP'$ satisfies $LP\leq(1+\epsilon)LP'$. The two approximation factors arise from the separation oracle and the binary-search accuracy, each contributing a factor of $\sqrt{1+\epsilon}$. For $m=O(1)$, separation is exact, so standard separation and optimization yield $LP'=LP$ in polynomial time.

We now describe the algorithm. It solves the restricted linear program, whose optimum is $LP'$, and for every agent $i$ it independently chooses one triple $(a,d,t^i)$ with probability $\frac{x^i_{a,d,t^i}}{8}$, where the $x^i_{a,d,t^i}$ are the values of this solution, and ignores agent $i$ with the remaining probability. It considers the chosen agents with their corresponding action in increasing order of $d$,
breaking ties arbitrarily, and when a triple $(a,d,t^i)$ is reached it incentivizes agent $i$ to choose $a$ with
the contract $t^i$ if the used budget is at most $d$, and skips the agent otherwise. We denote
its expected additional value by $ALG$.

\begin{lemma}
\label{lem:rounding}
    The algorithm is ex-post budget feasible, and every chosen agent is incentivized with probability
    at least $\frac12$.
\end{lemma}
\begin{proof}
    When the triple $(a,d,t^i)$ is reached, the algorithm either skips agent $i$, in which case no payment is made, or the used budget is at most $d$, so the remaining budget is at least $1-d$. As $t^i$ has deadline $d$ it pays at most $1-d$ on every outcome, so in both cases the contract is feasible. 
    
    Fix a triple $(a,d,t^i)$ with $d>0$ and condition on it being selected in the random sampling step, independently of the selections made for the other agents. Every agent preceding it in the algorithm had
    $d' \leq d$, and received no payment if it was skipped. Thus, as the solution of the restricted linear program satisfies the second constraint, and as every triple is chosen with probability $\frac18$ of its value in the LP, the expected sum of the payments to the preceding chosen agents, each truncated by $\min\{2d,1\}$, is at most $\frac18 \cdot 4d = \frac d2$. We consider truncated payments only for the analysis, since their expected sum is bounded by the LP. As $\min\{2d,1\}\geq d$, truncation cannot reduce total payments exceeding $d$ below $d$. Denoting this sum by $Y$, Markov's inequality gives $P[Y \geq d] \leq \frac{\EX[Y]}{d} \leq \frac{d/2}{d} = \frac12$, so the agent is incentivized with probability at least $\frac12$.

    We are left with the case of $d=0$. Fix a triple $(a,0,t^i)$, and condition on it being selected in the random sampling step. Since the agents are considered
    in increasing order of $d$, every agent preceding it was chosen with $d=0$, and the agent is
    skipped only if one of them made a positive payment. Letting $N$ be the number of preceding
    chosen agents which realize an outcome with a positive payment, the third constraint gives
    $\EX[N] \leq \frac18 \cdot 2 = \frac14$, so $P[N \geq 1] \leq \EX[N] \leq \frac14$ and the
    agent is incentivized with probability at least $\frac{3}{4} > \frac{1}{2}$.
\end{proof}

We are now ready to prove our main theorem.

\begin{proof}[Proof of Theorem~\ref{thm:lp_rounding}]
    Every triple $(a,d,t^i)$ is chosen with probability $\frac{x^i_{a,d,t^i}}{8}$, and by Lemma
    \ref{lem:rounding} it is incentivized with probability at least $\frac12$ conditioned on
    being chosen, in which case its additional expected value is $v^i_a$. Therefore
    $ALG \geq \frac{1}{16}\sum_{i,a,d,t^i}v^i_ax^i_{a,d,t^i} = \frac{1}{16}LP'$, and since
    $LP \leq (1+\epsilon)LP'$, Lemma \ref{lem:lp_upper_bound} gives
    $ALG \geq \frac{OPT}{16(1+\epsilon)}$. For $m=O(1)$, the LP is solved exactly, so
$LP'=LP$, giving $ALG\geq OPT/16$.

    For the running time, $|D|$ is polynomial in the input size, and by Lemma \ref{lem:pricing}
    the separation is performed in polynomial time for every fixed $\epsilon>0$,
    so the ellipsoid method returns a restricted linear program with polynomially many variables
    in polynomial time. The rounding is also performed in polynomial time.
\end{proof}

\subsection{Dependence on the Number of Agents}
\label{sec:const}

We also consider how the guarantees depend on the number of agents. When both the number of agents and outcomes are constant, the optimal adaptive policy can be computed exactly.

\begin{observation}
\label{obs:const-n-m}
If $n,m=O(1)$, the optimal adaptive policy can be computed in polynomial time.
\end{observation}
\begin{proof}
A single adaptive policy can be represented by a tree of depth at most $n$, where each node specifies which remaining agent to approach and which action to incentivize, and its $m$ outgoing edges correspond to the possible outcomes of that agent. Thus, each tree has at most $1+m+\cdots+m^{n-1}=O(m^n)$ nodes.

We enumerate the agent and action specified at every node. Since each node has at most $n\ell$ possible choices, the number of such trees is at most $(n\ell)^{O(m^n)}$, which is polynomial in $\ell$ since $n,m=O(1)$.

For each tree, we can compute the best possible matching contracts simultaneously by a linear program, with the variables as the payments for each outcome for each node in the tree. The constraints enforce that the target action in each node is incentive compatible against all actions of the corresponding agent, and that the total payment in each path is feasible. Therefore, since there are polynomially many trees with corresponding LPs of polynomial size, the optimal policy can be computed in polynomial time.
\end{proof}

 When the number of outcomes $m$ is unrestricted, a simple algorithm can achieve a $\frac{1}{n}$-approximation regardless of the number of actions per agent $\ell$. We will later show that this result is tight when both $\ell$ and $m$ are unrestricted.

\begin{theorem}
\label{thm:1agent_approx}
There is a polynomial-time $\frac{1}{n}$-approximation to the optimal adaptive policy.
\end{theorem}
\begin{proof}
    For every agent $i$, we compute the maximum expected value obtainable by $i$ if it is the only agent. This is exactly the expected value of the best action of agent $i$ that can be incentivized with budget $B$, which we can compute in polynomial time using a linear program.

Since in expectation the optimal adaptive policy cannot obtain more value from any agent than it could obtain from that agent alone with the full budget $B$, $OPT$ is at most the sum of these $n$ values. Therefore, incentivizing only the agent with the largest expected value achieves at least $\frac{1}{n}OPT$.
\end{proof}

When the number of agents and actions is constant, we can obtain a stronger guarantee. 
\begin{theorem}
\label{thm:ptas}
For every fixed $n,\ell$ and every $\epsilon>0$, there is a polynomial-time $(1-\epsilon)$-approximation to the optimal adaptive policy.
\end{theorem}

\begin{proof}
We assume $B=1$ and $\epsilon\in(0,1)$, and let $\delta=\epsilon/(4n)$. As in the general approximation algorithm, we ignore actions with non-positive expected value.

For every subset $S$ of agents, we construct a table indexed by target expected values $q$ on a geometric grid with ratio $1+\delta$, ranging from half the smallest positive $v^i_a$ to $\sum_i\max_a v^i_a$. Each entry stores a policy achieving expected value at least $q$, together with the initial budget required to execute it. We process subsets in increasing order of size, starting with the empty policy. Since $n$ and $\ell$ are constant and $\delta=\epsilon/(4n)$, the grid has polynomial size for fixed $\epsilon$.

Fix $S$ and a target $q$. We first consider policies already constructed for strict subsets of $S$. We then enumerate the first agent $i\in S$ and the action $a$ to incentivize. After each outcome $o_j$, we may use a policy from the already computed table for $S\setminus\{i\}$. For each grid value $q'$, let $b_{q'}$ be the required budget of the corresponding policy. We also include the empty policy, with value and budget $0$.

We construct an LP whose variables are $b$, the required initial budget, $t^i_j$, the payment to agent $i$ upon outcome $o_j$, and $x_{j,q'}$, indicating that after $o_j$ we use the policy corresponding to $q'$.

Since the LP has at most $\ell$ constraints coupling different outcomes, we will show that it admits an optimal solution in which at most $\ell$ outcomes have fractional assignments. For these outcomes, we replace the fractional choices by the empty policy, without changing the payments. To bound the resulting loss, we ensure that each such outcome has probability at most $\delta/\ell$.

We therefore enumerate all possible grid values $q'$ for each outcome satisfying $p^i_{a,j}>\delta/\ell$, fixing $x_{j,q'}=1$ for the chosen value. Since the probabilities sum to $1$, there are at most $\ell/\delta$ such outcomes. Thus, the number of choices is polynomial for fixed $\ell,\delta$.

Since the LP allows fractional assignments, its budget constraint may hold even when a continuation policy assigned positive weight requires more budget than $b$. To prevent this, we enumerate an upper bound $\beta$ on the budget required by any continuation policy. Specifically, we consider each budget $b_{q'}$ stored in the table for $S\setminus\{i\}$, including $0$ for the empty policy, as a possible value of $\beta$. For each choice, we require $\beta\leq b$ and exclude all continuation policies with $b_{q'}>\beta$. This does not exclude any integral solution, since we can choose $\beta$ as the largest budget required by its continuation policies.

Finally, to bound the rounding loss, we consider only grid values $q'\leq q/(1-\delta)$. Any policy with a larger guaranteed value can instead be used directly, without approaching agent $i$, and was already considered among the policies for strict subsets of $S$.

For each choice of $i$, $a$, $\beta$, and the target values $q'$ assigned to the high-probability outcomes, we solve the following LP:
\begingroup
\setlength{\jot}{1pt}
\begin{align*}
    \min\quad & b\\
    \text{s.t.}\quad
    & \sum_{q'}x_{j,q'}=1 \qquad \forall j\\
    & t^i_j+\sum_{q'}b_{q'}x_{j,q'}\leq b
      \quad \forall j\\
    & \sum_j(p^i_{a,j}-p^i_{a',j})t^i_j
      \geq c^i_a-c^i_{a'}
      \quad \forall a'\neq a\\
    & v^i_a+\sum_{j,q'}p^i_{a,j}q'x_{j,q'}
      \geq \tfrac{q}{1-\delta}\\
    & \beta\leq b\\
    & x_{j,q'}=0 \quad
      \substack{\forall j,q':\ b_{q'}>\beta\\
      \text{or }q'>\tfrac{q}{1-\delta}}\\
    & x_{j,q'}=1
      \quad \text{for enumerated choices}\\
    & t^i_j,x_{j,q'}\geq0 \quad \forall j,q'.
\end{align*}
\endgroup

The first two constraints allow fractional assignments after each outcome and ensure that the payment plus the weighted average of the continuation budgets is at most $b$. The next two enforce incentive compatibility and an expected value of at least $q/(1-\delta)$, allowing for the loss incurred by rounding. The remaining constraints implement the choices described above.

We now show that the LP has an optimal solution in which at most $\ell$ outcomes $j$ have $0<x_{j,q'}<1$ for some $q'$. Fix an optimal value of $b$. 
Since the LP excludes all continuation values $q'$ with $b_{q'}>\beta$, and requires $\beta\leq b$, its constraints imply that $\sum_{q'}x_{j,q'}=1$ and $0\leq t^i_j\leq\sum_{q'}x_{j,q'}(b-b_{q'})$. Thus, for each outcome $j$, these variables can be expressed as a convex combination of assignments, each setting exactly one $x_{j,q'}=1$, all other assignment variables to $0$, and the payment $t^i_j$ to either $0$ or $b-b_{q'}$. We refer to the weights of these assignments as coefficients.

The coefficients of these combinations satisfy $m$
constraints requiring them to sum to $1$ for each outcome, and at most $\ell$ additional constraints from IC and
the expected-value requirement. If more than $m+\ell$ coefficients are positive, their
contributions to these constraints are linearly dependent. We can therefore adjust the coefficients while preserving their sums for each outcome, the values of the IC and expected-value expressions, and the value of \(b\), until one becomes zero.
Repeating this gives an optimal solution with at most
$m+\ell$ positive coefficients.  
Since the coefficients for each outcome sum to $1$, every outcome requires at least one positive coefficient. Moreover, if $0<x_{j,q'}<1$ for some $q'$, at least two coefficients must be positive. Therefore, at most $\ell$ outcomes have fractional variables, as claimed.

By our enumeration, every outcome requiring rounding satisfies $p^i_{a,j}\leq\delta/\ell$. We replace its choice by the empty policy, keeping the payment unchanged. This preserves incentive compatibility and ex-post feasibility. Since every allowed grid value satisfies $q'\leq q/(1-\delta)$, the total loss in guaranteed expected value is at most $\ell\cdot\frac{\delta}{\ell}\cdot
    \frac{q}{1-\delta}
    =\frac{\delta q}{1-\delta}$.
    
Therefore, the resulting policy achieves expected value at least
$\frac{q}{1-\delta}-\frac{\delta q}{1-\delta}=q$,
without increasing its required budget. For each $S$ and $q$, we store the policy requiring the
smallest budget among all choices of $i$, $a$, $\beta$,
and the enumerated assignments, as well as the previously
computed policies for strict subsets of $S$ that achieve $q$. 

The argument above shows that every policy constructed by the LP achieves its target value $q$. It remains to show that the table contains a policy whose target value approximates that of an optimal adaptive policy.  We establish this guarantee by induction on $|S|$.

\begin{claim}
\label{clm:ptas_induction}
For every subset $S$ of agents and every initial budget $b$,
the table for $S$ contains a policy requiring budget at most $b$,
whose target value is at least
$\left(\frac{1-\delta}{1+\delta}\right)^{|S|}$
times the value obtainable from $S$ with budget $b$
by an optimal adaptive policy.
\end{claim}
\begin{proof}
    We prove the claim by induction on $|S|$. The claim is immediate for $S=\emptyset$. Otherwise, consider an optimal policy for $S$. By induction, its policy after each outcome can be replaced by an entry from the table for the remaining agents, without increasing the required budget. Each continuation uses at most $|S|-1$ agents, so by the induction hypothesis, its replacement guarantees at least $(\frac{1-\delta}{1+\delta})^{|S|-1}$ times its original expected value. Since $v^i_a\geq0$, the same bound holds for the total expected value.

Choose the largest grid value $q$ not exceeding $(1-\delta)$ times this guarantee. If one of the chosen policies for the remaining agents has target value greater than $q/(1-\delta)$, that policy can be used directly, achieving at least $q$ with no greater budget. Otherwise, these choices form a feasible integral solution to one of our LPs with no greater initial budget. By the rounding argument, we obtain a policy achieving at least $q$. Since consecutive grid values differ by a factor of $1+\delta$,
we have $q\geq\frac{1-\delta}{1+\delta}V$ where $V$ is the guarantee obtained using the
continuation policies from the previous tables.
By the induction hypothesis, this is at least
$\left(\frac{1-\delta}{1+\delta}\right)^{|S|}$
times the value of the optimal adaptive policy.
This completes the induction.
\end{proof}

Finally, since $\delta=\epsilon/(4n)$, we get  $\left(\frac{1-\delta}{1+\delta}\right)^n
    \geq 1-\epsilon$. 
    
    There are $2^n$ subsets, polynomially many grid values, and polynomially many choices to enumerate for fixed $n,\ell,\epsilon$. Each LP has polynomial size, completing the proof.
\end{proof}

\subsection{Computational Hardness}
\label{sec:hardness}

The preceding algorithm provides a PTAS for fixed
$n$ and $\ell$. Computing an optimal policy remains
NP-hard even when $n=\ell=2$.

\begin{theorem}
\label{thm:2agent_hard}
Computing an optimal adaptive policy is NP-hard, even for
two agents with two actions each.
\end{theorem}
\begin{proof}
    We reduce from Subset Sum. Given positive integers
$a_1,\ldots,a_M$ and a target $T$, let
$A=\sum_{j=1}^M a_j$, where $0<T<A$.

Consider two agents with budget $B=1$ and $M+1$ outcomes.
The outcome values are $r_1=0$ and $r_j=1$ for every
$j\geq2$. Each agent has two actions, where Action $1$ for both agents $i$
has cost $c^i_1=0$ and produces $o_1$ with probability $1$.

For Agent $1$, Action $2$ has cost $T/A$ and produces
$o_{j+1}$ with probability $a_j/A$ for every $j\in[M]$.
For Agent $2$, Action $2$ has cost $1$ and produces
$o_2$ with probability $1$.

To incentivize Agent $1$ to choose Action $2$, the
contract must satisfy $\sum_{j=1}^M a_j t^1_{j+1}\geq T+At^1_1$.
We can assume $t^1_1=0$, since this payment only
makes the IC constraint harder to satisfy.
As every payment is at most $1$, the set
$J=\{j\in[M]:t^1_{j+1}>0\}$ must satisfy $\sum_{j\in J}a_j\geq T$.
Therefore, paying $1$ on the outcomes corresponding
to any such set $J$ incentivizes Action $2$.

Agent $2$ can be incentivized to choose Action $2$ only
when the remaining budget is $1$, since its IC constraint
requires $t^2_2-t^2_1\geq1$.
Therefore, if agent $1$ acts first, agent $2$ can be
incentivized if and only if agent $1$ makes no positive
payment. This occurs with probability
$1-\frac{1}{A}\sum_{j\in J}a_j$. Incentivizing agent $2$ first yields value exactly $1$, and uses the entire budget, so Agent $1$ cannot be incentivized afterwards.
Since each incentivized agent contributes exactly $1$ to the expected (and also actual) value, the optimal value is $OPT=2-\frac{1}{A}
    \min_{J\subseteq[M]:\,\sum_{j\in J}a_j\geq T}
    \sum_{j\in J}a_j$.

Therefore, $OPT=2-T/A$ if and only if there exists
a subset $J\subseteq[M]$ satisfying
$\sum_{j\in J}a_j=T$.
Computing an optimal adaptive policy is therefore NP-hard. 
\end{proof}

Fixing the number of outcomes does not restore
tractability when the number of agents is arbitrary.

\begin{theorem}
\label{thm:binary_hard}
Computing the optimal adaptive policy is NP-hard,
even for binary-action agents with binary outcomes.
\end{theorem}
\begin{proof}
    We reduce from Knapsack with integer weights $w_i$,
values $v_i$, and capacity $W$. Set $B=1$ and
construct one agent per item with
$$
p^i_1=(1,0),\quad p^i_2=(0,1),\quad c_i=w_i/W,\quad r^i=(0,v_i).
$$
Incentivizing agent $i$ requires
$t^i_2-t^i_1\geq w_i/W$, incurs a deterministic
payment of $\ge w_i/W$, and adds expected
value $v_i$. So every realized set of
incentivized agents is a feasible Knapsack
solution. However, every feasible Knapsack
solution is implementable using $t^i=(0,w_i/W)$.
Thus the optimal value equals
the Knapsack optimum.
\end{proof}

We finally turn to approximation hardness when there is an arbitrary number of actions and outcomes. We show that the $1/n$-approximation is
tight, even for identical agents and any fixed $n$.

\begin{theorem}
\label{thm:fixed_n_hardness}
Fix $n\geq2$. For every constant $\alpha\in(\frac{1}{n},1]$, there is no polynomial-time $\alpha$-approximation to the optimal adaptive policy unless $P=NP$, even for identical agents.
\end{theorem}
\begin{proof}
Fix $\alpha\in(\frac{1}{n},1]$. We use the hardness result of \cite[Section~1, "Location of the gap"]{DGKR03}. For every integer $K\geq 3$ and every $\epsilon>0$, it is NP-hard to distinguish between $K$-uniform hypergraphs $H=(V,E)$ for which there exists a vertex cover of size at most $\left(\frac{1}{K-1}+\epsilon\right)|V|$, 
and those for which every vertex cover has size at least $(1-\epsilon)|V|$. 

Equivalently, let $U=E$ and create one set
$S_j$ for every vertex $j\in V$, containing
the hyperedges incident to that vertex.
Writing $M=|V|$ and denoting the minimum
cover size by $\tau$, it is NP-hard to
distinguish between Set Cover instances satisfying $\tau\leq \left(\frac{1}{K-1}+\epsilon\right)M$
and those satisfying $\tau\geq(1-\epsilon)M$.

Given such a Set Cover instance, let
$w=\frac{1}{M+1}$ and consider $n$ identical
agents with $M+2$ outcomes. For every
$e\in U$, each agent has an action $a_e$.
The instance is specified
in the following table.

\begin{table}[h]
  \centering
  \begin{tabular}{l|ccccc|c}
     & $o_1$ & $o_2$ & $\dots$ & $o_{M+1}$ & $o_{M+2}$ & $c$ \\
    \hline
    Action $a_e$ & $w(1+|\{j : e \in S_j\}|)$ & \multicolumn{3}{c}{$w$ if $e \notin S_{j-1}$, else $0$} & $0$ & $0$ \\
    Action $1$ & $0$ & $w$ & $\dots$ & $w$ & $w$ & $2w$\\
    Action $2$ & $w$ & $w$ & $\dots$ & $w$ & $0$ & $w$ \\
    \hline
    $r_j$ & $0$ & $0$ & $\dots$ & $0$ & $\frac{1}{w}$ &\\
  \end{tabular}%
  \par\vspace{-6pt}
\end{table}

Only Action $1$ has positive expected value,
equal to $1$. Consider a contract $t$ which incentivizes Action $1$. From the IC constraint with Action $2$ we get $w(t_{M+2}-t_1) \geq w$, so as every payment is at most $1$ we have $t_{M+2} = 1$ and $t_1 = 0$. From the IC constraint with Action $a_e$ we have $w(t_{M+2} + \sum_{j : e \in S_j}t_{j+1}) \geq 2w$, so
    $\sum_{j : e \in S_j}t_{j+1} \geq 1$ for every $e \in U$. Hence $\{S_j : j\in[M],\,t_{j+1}>0\}$ is a set cover,
   and $t$ must pay a positive amount on at least $\tau$ of the outcomes $o_2, \dots, o_{M+1}$. Therefore, a payment of $1$ for outcome $o_{M+2}$ and the outcomes corresponding to a minimum set cover incentivizes Action $1$.

Since under Action $1$, each outcome $o_2,\ldots,o_{M+2}$ is realized with probability $\frac{1}{M+1}$, the minimum probability that an incentivized agent makes a positive payment is $\frac{\tau+1}{M+1}$. After any positive payment, the remaining budget is strictly smaller than $1$, while incentivizing Action $1$ requires a payment of $1$ on $o_{M+2}$. Thus, no additional agent can be incentivized to choose Action $1$ after any previous agent is paid a nonzero amount. If no payment is made, the full budget remains. 
Since no other action yields positive expected value, an optimal policy repeatedly incentivizes Action $1$ for different agents in some order (the order does not matter as the agents are identical), using a contract minimizing the probability of a positive payment. Therefore, $OPT=
\sum_{j=0}^{n-1}
\left(1-\frac{\tau+1}{M+1}\right)^j$.

We now bound $OPT$ using the Set Cover hardness gap, distinguishing between the case $\tau\leq(\frac{1}{K-1}+\epsilon)M$ and the case $\tau\geq(1-\epsilon)M$. In the first case, $\frac{\tau+1}{M+1}
\leq
\frac{\left(\frac{1}{K-1}+\epsilon\right)M+1}{M+1}$, 
which approaches $\frac{1}{K-1}+\epsilon$. Thus, by choosing  a sufficiently large $K$ and a sufficiently small $\epsilon$, $OPT$ can be made arbitrarily close to $n$. In the second case, $\frac{\tau+1}{M+1}\geq 1-\epsilon$,
so $OPT$ can be made arbitrarily close to $1$.

Since $\alpha>\frac{1}{n}$, we have $\alpha n>1$. We can therefore choose a sufficiently large $K$ and a sufficiently small $\eps$ so that an $\alpha$-approximation has a strictly larger value in the first case than the optimum in the second case. Since $n$ is constant, the expected value of the
returned policy can be computed in polynomial time
by enumerating all possible outcome histories. Such an algorithm would distinguish the two NP-hard cases, a contradiction unless $P=NP$.
\end{proof}

We obtain the following inapproximability result:

\begin{corollary}
\label{cor:general_hardness}
When the numbers of agents, actions, and outcomes are unrestricted, no polynomial-time constant-factor approximation to the optimal adaptive policy exists unless $P=NP$, even for identical agents.
\end{corollary}
\begin{proof}
Fix any constant $\alpha \in (0,1)$, and choose a
fixed integer $n>1/\alpha$. By the preceding
theorem, no polynomial-time
$\alpha$-approximation exists for this $n$.
\end{proof}

\section{Simple Contracts}
\label{sec:simple}
Our general algorithm chooses an agent's contract based on the full instance. We next ask whether substantially simpler contract
structures can still provide good approximation guarantees.

\subsection{The Value of Contract Adaptivity}

We first ask whether adapting contracts to previously observed outcomes is
actually useful. We compare the optimal fully adaptive policy with a static
policy, which fixes the contract for each agent in advance but may still adapt
the order in which agents are approached. 

\begin{claim}
There is an instance for which
$$
    OPT_{\mathrm{static}} \leq \frac{5}{6}OPT.
$$
\end{claim}
\begin{proof}
Consider an instance with budget $B=1$ and three risk-free agents, each with three outcomes. We may assume $t^i_1=0$, and therefore write a contract for agent $i$ as $(t^i_2,t^i_3)$. The instance is defined in Table~\ref{tab:risk-free-gap}.

\begin{table}[h]
\centering
\begin{tabular}{c|ccc|cc|c}
 & $p^i_{2,1}$ & $p^i_{2,2}$ & $p^i_{2,3}$ 
 & $r^i_2$ & $r^i_3$ & $c_i$ \\
\hline
Agent $1$ & $0$ & $\frac{1}{3}$ & $\frac{2}{3}$ 
          & $\frac{2}{3}$ & $\frac{2}{3}$ & $\frac{1}{6}$ \\
Agent $2$ & $0$ & $\frac{1}{2}$ & $\frac{1}{2}$ 
          & $1$ & $1$ & $\frac{1}{2}$ \\
Agent $3$ & $0$ & $1$ & $0$ 
          & $1$ & $0$ & $1$ \\
\end{tabular}
\caption{The effort distributions, rewards, and costs in the lower-bound instance.}
\label{tab:risk-free-gap}
\end{table}

For Agent $1$, exerting effort results in $o_2$ with probability $\frac13$ and in $o_3$ with probability $\frac{2}{3}$, with
$$
    r^1_2=r^1_3=\frac{2}{3}
    \qquad\text{and}\qquad
    c_1=\frac{1}{6}.
$$
Thus, incentivizing Agent $1$ requires
$$
    t^1_2+2t^1_3\geq\frac{1}{2}.
$$
For Agent $2$, exerting effort again results in $o_2$ and $o_3$ with probability $\frac12$ each, with
$$
    r^2_2=r^2_3=1
    \qquad\text{and}\qquad
    c_2=\frac12,
$$
so its incentive constraint is
$$
    t^2_2+t^2_3\geq1.
$$
Finally, for Agent $3$, exerting effort deterministically results in $o_2$, with
$$
    r^3_2=1
    \qquad\text{and}\qquad
    c_3=1.
$$
Hence every contract incentivizing Agent $3$ satisfies $t^3_2\geq1$, and Agent $3$ can therefore be incentivized only while the full budget remains.

We first describe an adaptive policy. Incentivize Agent $1$ using the contract
$(\frac{1}{2},0)$. If Agent $1$ realizes $o_2$, the remaining budget is $\frac{1}{2}$, and we incentivize Agent $2$ using $(\frac{1}{2},\frac{1}{2})$. The total value in this case is $\frac{5}{3}$.

If Agent $1$ realizes $o_3$, no payment is made and the full budget remains. We then incentivize Agent $2$ using $(0,1)$. If Agent $2$ realizes $o_2$, no payment is made and Agent $3$ can also be incentivized. Otherwise, agent $2$ realizes $o_3$, and the budget is exhausted. Thus, conditioned on Agent $1$ realizing $o_3$, the expected value is
$$
    \frac{2}{3}+1+\frac{1}{2}=\frac{13}{6}.
$$
Therefore, the expected value of this policy is
$$
    \frac{1}{3}\cdot\frac{5}{3}+
    \frac{2}{3}\cdot\frac{13}{6}
    =2.
$$

We next upper bound the value of any static policy. Fix the contracts offered to all three agents, while allowing the principal to adaptively choose their order.

If Agent $2$ is not incentivized, the expected value is at most $\frac{2}{3}+1=\frac{5}{3}$, so assume its contract incentivizes it.

Suppose first that $t^2_2,t^2_3>0$. Agent $2$ then makes a positive payment with probability $1$. Therefore, Agents $2$ and $3$ cannot both be incentivized on the same realization: if Agent $2$ is incentivized first, less than the full budget remains afterwards, while if Agent $3$ is incentivized first, the entire budget is exhausted. Hence the expected value is at most
$$
    \frac{2}{3}+1=\frac{5}{3}.
$$

Suppose instead that Agent $2$ is paid on only one of $o_2,o_3$. Its incentive constraint then implies that this payment is $1$, so Agent $2$ can be incentivized only while the full budget remains. The same is true of Agent $3$.

If Agent $1$ is incentivized first, either both payments in its contract are positive, in which case neither Agent $2$ nor Agent $3$ can subsequently be incentivized, or one payment is zero. By the incentive constraint
$$
    t^1_2+2t^1_3\geq\frac{1}{2},
$$
the probability that Agent $1$ makes no payment is at most $\frac{2}{3}$. Conditioned on making no payment, the maximum expected value obtainable from Agents $2$ and $3$ is
$$
    1+\frac{1}{2}\cdot1=\frac{3}{2},
$$
attained by incentivizing Agent $2$ first and Agent $3$ if Agent $2$ makes no payment. Thus, the expected value is at most
$$
    \frac{2}{3}+\frac{2}{3}\cdot\frac{3}{2}
    =\frac{5}{3}.
$$

If Agent $2$ is incentivized first, it makes no payment with probability $\frac{1}{2}$. Conditioned on this event, the maximum expected value obtainable from Agents $1$ and $3$ is at most
$$
    \frac{2}{3}+\frac{2}{3}\cdot1
    =\frac{4}{3},
$$
since Agent $3$ can be incentivized after Agent $1$ only if Agent $1$ makes no payment, which occurs with probability at most $\frac{2}{3}$. Therefore, the expected value is at most
$$
    1+\frac{1}{2}\cdot\frac{4}{3}
    =\frac{5}{3}.
$$

Finally, incentivizing Agent $3$ first exhausts the entire budget and gives value $1$. Thus, every static policy has expected value at most $\frac{5}{3}$.

This value is attained by offering $(\frac{1}{2},0)$ to Agent $1$, $(\frac{1}{2},\frac{1}{2})$ to Agent $2$, and $(1,0)$ to Agent $3$. Incentivize Agent $1$ first; if it realizes $o_2$, incentivize Agent $2$, and if it realizes $o_3$, incentivize Agent $3$. Therefore,
$$
    OPT_{\mathrm{static}}=\frac{5}{3}<2\leq OPT,
$$
and hence
$$
    OPT_{\mathrm{static}}\leq\frac{5}{6}OPT.
$$
\end{proof}

Thus, contract adaptivity can strictly improve the principal's value, even for
risk-free agents. We next ask whether the fixed contracts in a static policy
can themselves be chosen using only local information about each agent.

\subsection{Local Contracts}

Static policies may still use the entire instance when choosing the contract
assigned to each agent. We now impose the stronger local restriction, requiring contracts to be chosen by a common deterministic rule that is independent of the instance and uses only each agent's own parameters and the common budget. An example of such a local contract is the \emph{minimax contract}, which is the contract that minimizes $\max_j t_j$ among all contracts that incentivize effort.

We show this restriction can be arbitrarily bad, even when all agents are identical.

\begin{claim}
\label{clm:identical_local}
For every $\epsilon>0$, there exists an instance with identical agents such that every local policy achieves expected value less than $\epsilon OPT$.
\end{claim}
\begin{proof}
    Let $\delta, \omega >0$ be arbitrarily small. Consider $n$ identical agents, each with $3$ outcomes with $r_1=r_2=0$ and $r_3=\frac{1}{\delta}$. The budget is $B=1$.

    \begin{table}[h]
  \centering
  \begin{tabular}{l|ccc|c}
     & $o_1$ & $o_2$ & $o_3$ & $c$ \\
    \hline
    
    Action $1$ & $\frac{1}{2}+\delta$ & $\frac{1}{2}-\delta$ & $0$ & $0$\\
    Action $2$  & $\frac{1}{2}-\delta$ & $\frac{1}{2}$ & $\delta$ & $(1+\omega)\delta$ \\
    \hline
    $r_j$ & $0$ & $0$ & $\frac{1}{\delta}$ &\\
  \end{tabular}
  \caption{IC requires $t_2+t_3 \geq 1+\omega$.}
\end{table}

    Action $1$ has a cost of $0$ and leads to Outcome $o_1$ with probability $\frac{1}{2}+\delta$, and Outcome $o_2$ with probability $\frac{1}{2}-\delta$. Action $2$ has a cost of $(1+\omega)\delta$, and leads to $o_1$ with probability $\frac{1}{2}-\delta$, $o_2$ with probability $\frac{1}{2}$, and $o_3$ with probability $\delta$.

    As $o_3$ is the only outcome with positive reward, and only Action $2$ leads to $o_3$, it is sufficient to consider contracts which incentivize agent $i$ to choose Action $2$. Let $t_2, t_3$ be the payments to $o_2,o_3$ respectively. In order to satisfy IC, we have:
    \begin{align*}
    \frac{1}{2}t_2 + \delta t_3 - (1+\omega)\delta &\geq  (\frac{1}{2}-\delta)t_2\\
    t_2 +t_3 &\geq  1+\omega.
    \end{align*}
    
        Note that an agent incentivized to choose Action $2$ achieves $o_3$ with probability $\delta$, for an expected value of $\delta \cdot \frac{1}{\delta} = 1$, while an agent choosing Action $1$ has an expected value of $0$. Thus the expected value of an algorithm is exactly the expected number of agents it incentivizes. Once an agent realizes outcome $o_2$ or $o_3$, no additional agents can be incentivized by the same contract, as $t_2+t_3 \geq 1+\omega >1$, so after paying either outcome the remaining budget is insufficient to pay the other outcome for the next agent. As one of these outcomes occurs at a total probability of $\frac{1}{2}+\delta$ per agent, the expected number of incentivized agents is at most $\frac{1}{\frac{1}{2}+\delta} < 2$, for an expected value of less than $2$.

   For a lower bound on $OPT$, consider the adaptive policy that, whenever the remaining budget is $b$, uses
$t_2=1+\omega-b$ and $t_3=b$. We choose $n,\delta, \omega$ such that $n \geq 2\log_2 \frac{1}{\omega}$ and $\delta \leq \log_2^{-2} \frac{1}{\omega}$. The contract is feasible while $b \geq \frac{1+\omega}{2}$, so the total paid for $o_2$ may not exceed $\frac{1-\omega}{2}$. Since the payment for $o_2$ starts at $\omega$ and doubles after every agent with a realized outcome of $o_2$, the budget suffices for $o_2$ to be realized $\log_2 (\frac{1-\omega}{2\omega}+1)$ times. Each agent realizes outcome $o_2$ or $o_3$ with probability $\frac{1}{2}+\delta$, and otherwise achieves $o_1$ and is paid nothing, leaving the budget unchanged, so more than $\frac{3}{2}$ agents are incentivized in expectation for each realization of $o_2$ or $o_3$. Conditioned on a realized outcome of $o_2$ or $o_3$, the realized outcome is $o_3$ with probability $\frac{\delta}{\frac{1}{2}+\delta} < 2\delta$, so $o_3$ is realized by an agent before $o_2$ is realized $\log_2 (\frac{1-\omega}{2\omega}+1)$ times with probability at most $2\delta\log_2 (\frac{1-\omega}{2\omega}+1) = o(1)$. Thus, with probability $1-o(1)$, either $o_2$ is realized
$\log_2 (\frac{1-\omega}{2\omega}+1)$ times, or all $n$ agents are incentivized.
Since $n \geq 2\log_2 \frac{1}{\omega}$, the expected value of this algorithm is
$\Omega(\log \frac{1}{\omega})$.
\end{proof}

Since the agents are identical, every local policy assigns the same contract
to every agent. Thus, the claim rules out all local contract rules in this
setting, including the rule assigning each agent its minimax contract.

\subsubsection{Structure of Optimal Contracts}
In search of positive results for local contracts, we consider risk-free agents. Recall that risk-free agents have two actions, and shirking deterministically leads to outcome $o_1$. The motivation for considering risk-free agents is settings in which valuable outcomes cannot occur without effort.

Even in this setting, there is no universal optimal payment pattern. In particular, in certain cases an optimal policy may need to spread payments across several outcomes, while in other cases it must concentrate the payment on a single outcome.

\begin{example}
\label{ob:balanced}
There is an instance in which every optimal policy pays on both nonzero outcomes of an agent, even though the agent can be incentivized by paying on only one outcome.
\end{example}
\begin{proof}
Consider two identical risk-free agents and budget $B=2$. For each agent, exerting effort results in $o_2$ and $o_3$ with probability $\frac12$ each, with
\[
    r^i_2=r^i_3=1
    \qquad\text{and}\qquad
    c_i=1.
\]
Thus, a contract incentivizing effort must satisfy
\[
    t^i_2+t^i_3\geq2.
\]

By offering $(1,1)$ to each agent, the principal can incentivize both agents for every realization and obtain value $2$. In order to guarantee that the second agent can also be incentivized, the payment made to the first agent can be at most $1$ on either outcome. Together with the incentive constraint, this requires
\[
    t^i_2=t^i_3=1.
\]
Hence an optimal policy must spread the payment across both nonzero outcomes.
\end{proof}

\begin{example}
There is an instance in which an optimal policy must concentrate the payment on a single outcome.
\end{example}
\begin{proof}
Consider two risk-free agents and budget $B=2$. For Agent $1$, exerting effort results in $o_2$ and $o_3$ with probability $\frac12$ each, with
\[
    r^1_2=r^1_3=2
    \qquad\text{and}\qquad
    c_1=1.
\]
Thus, its incentive constraint is
\[
    t^1_2+t^1_3\geq2.
\]

For Agent $2$, exerting effort results in $o_2$ with probability $\frac12$ and in $o_1$ otherwise, with
\[
    r^2_2=1
    \qquad\text{and}\qquad
    c_2=1.
\]
Hence Agent $2$ can be incentivized only by a contract with $t^2_2\geq2$, and therefore only while the full budget remains.

Agent $1$ has expected value $2$ whenever it is incentivized. If both $t^1_2$ and $t^1_3$ are positive, some budget is spent with probability $1$, so Agent $2$ cannot subsequently be incentivized. The resulting expected value is therefore at most $2$.

Instead, incentivize Agent $1$ using $(2,0)$. With probability $\frac12$, Agent $1$ realizes $o_3$ and makes no payment, leaving the full budget available to incentivize Agent $2$. This gives expected value
\[
    2+\frac12\cdot\frac12
    =\frac94.
\]
Thus, in an optimal policy the payment to Agent $1$ is concentrated on a single outcome.
\end{proof}

However, for risk-free agents, which deterministically realize outcome $o_1$ when shirking, a local policy using minimax contracts achieves a constant-factor approximation.

\subsubsection{A Constant-Factor Local Policy}
\label{sec:risk_free_upper}
We now show that risk-free agents admit a local policy that is a constant-factor approximation to $OPT$. For each agent $i$, let
$$
p^i_{-1}=\sum_{j\neq1}p^i_{2,j}.
$$

Recall that $v_i=\sum_jp^i_{2,j}r^i_j$ is the principal's additional expected value from agent $i$ exerting effort. We discard agents with $v_i\leq0$. Agents with $c_i=0$ can be incentivized for free, so we add their value to every solution and remove them. Agents with $c_i>0$ and $p^i_{-1}=0$ cannot be incentivized and can be discarded. Thus, for the remaining agents, $c_i>0$ and $p^i_{-1}>0$.
We use the contract
$$
t^i_1=0,
\qquad
t^i_j=\frac{c_i}{p^i_{-1}}
\quad\text{for every }j\neq1.
$$
This is the minimax contract for agent i. Since the same contract-selection rule is applied to every agent, using only its own parameters and the common budget, the resulting contract assignment is local. We assume w.l.o.g., that $\frac{c_i}{p^i_{-1}} \leq B$ for every agent, as otherwise agent $i$ cannot be incentivized by any algorithm.
\paragraph{Warm-up.} As a warm-up, we begin with the case where for every agent $i$ $p^i_{2,1}=0$. Since risk-free agents satisfy $p^i_{1,1}=1$ for every $i$, an optimal contract will never pay a shirking agent. This simplifies the problem, as by setting $t^i_j=c_i$ for every outcome $j\neq 1$, we guarantee that every incentivized agent receives a payment of exactly $c_i$, and provides expected additional value $v_i$. Thus, for this static choice of payment the resulting problem is a $0$-$1$ knapsack problem.

\begin{theorem}[Special case]
\label{thm:p+q=1}
If $p^i_{2,1}=0$ for every agent i, then the local minimax contracts admit a policy achieving expected value at least $\frac{1}{2}OPT$.
\end{theorem}
\begin{proof}
    Fix an optimal adaptive algorithm for maximizing value under budget $B$. Recall that $OPT$ is the expected value of the principal over the agent realizations in this algorithm. Let $S_{OPT}$ be the distribution over the sets of agents incentivized by this optimal algorithm.

    Consider the problem with static payments, so for every agent $i$ $t^i_1=0$ and $t^i_j=c_i$ for every $j \neq 1$. Note that this payment minimizes the maximum ex-post payment of the principal for agent $i$ over all alternative payments that incentivize $i$. Thus, since the budget constraint needs to be satisfied ex-post, for any remaining budget $b$ if there exists any budget-feasible contract that can afford to incentivize $i$, this contract is a budget-feasible contract for incentivizing agent $i$ to exert effort.

    Let $x_i \in [0,1]$ be the probability agent $i$ is incentivized in the optimal adaptive algorithm, i.e. $x_i = \Pr_{S \sim S_{OPT}}[i \in S]$. As the expected payment to each incentivized agent is at least $c_i$, the expected value is $OPT \leq \sum_i x_iv_i$. Additionally, since the expected cost must be under budget, $\sum_i x_ic_i \leq B$.

    Using this definition, we can provide an upper bound on the expected value of $OPT$ by using the fractional knapsack problem. Formally, we get $OPT \leq \max \{\sum_i y_i v_i| \sum_i y_ic_i \leq B, y_i \in [0,1]\}$. As it is known that the optimal integer knapsack solution achieves at least half the fractional knapsack optimum, we get $OPT \leq 2\max \{\sum_i y_i v_i| \sum_i y_ic_i \leq B, y_i \in \{0,1\}\}$.

    Since $p^i_{2,1}=0$ for every agent i, the integer knapsack solution is achievable by a local policy that sets $t^i_1=0$ and $t^i_j=c_i$ for every $j\neq 1$. Therefore, this policy achieves at least $\frac{1}{2}OPT$.
\end{proof}

\paragraph{General Case.} The restriction to the case where $p^i_{2,1}=0$ for every agent $i$ enabled us to remove the stochastic aspect of the payments, since effort always produces a nonzero outcome. In general, even for risk-free agents, this is not the case. However, we can still achieve a constant approximation ratio. Our proof uses ideas from \cite{DGV08}, but requires significant modification in order to handle the ex-post feasibility constraints and the adaptive payments.

    For simplicity, w.l.o.g. assume $B=1$. Given some $\sigma \in (0,1)$, we  partition the set of agents into light and heavy agents, denoted $N_L$ and $N_H$, respectively. An agent $i$ is {\em light} (i.e., $i \in N_L$) if $\frac{c_i}{\sum_{j \neq 1} p^i_{2,j}} \leq \sigma$. Otherwise, an agent $i$ is {\em heavy} (i.e., $i \in N_H$).  Intuitively, an agent is light if it can be incentivized with a remaining budget of at most $\sigma$. Assume the light agents are $\{1, 2, 3, \dots , |N_L|\}$, and they are ordered such that $\frac{v_1}{c_1} \geq \frac{v_2}{c_2} \geq \dots$.

    Our algorithm will make use of several useful benchmark values. Let $m_1 = \max_{i \in N_H} \frac{v_i}{p^i_{-1}}$ be the maximum expected value across all heavy agents, conditioned on the agent achieving an outcome other than $o_1$. We denote by $S_{success}$ a distribution over subsets $S \subseteq [n]$, where each $S$ represents the set of agents which, assuming they exert effort, realize an outcome other than $o_1$. $m_1^{success}$ is the expectation over $S_{success}$ of the expected value of the highest value heavy agent, i.e. $m_1^{success} = \EX_{S \sim S_{success}}(\max_{i \in S\cap N_H} \frac{v_i}{p^i_{-1}})$. Note that both $m_1$ and $m^{success}_1$ can be computed in polynomial time. All maxima over empty sets in this subsection are defined as $0$.
    
    Next, we define $m_G$, which serves as a bound for a greedy choice of the light items. In order to define $m_G$, we first define $M_k = \sum_{i=1}^k c_i$ to be the expected total payment to the first $k$ light agents, with $M_0=0$. Let $n^*\leq |N_L|$ be the maximum number of agents for which $M_{n^*-1} \leq 1-\sigma$, or $n^*=0$ if there are no light agents. Note that for every light agent $i$, $c_i \leq \frac{c_i}{p^i_{-1}} \leq \sigma$, so if there is at least one light agent then $M_{n^*} \leq 1$, $M_{n^*} \geq 1- \sigma$ unless $n^* = |N_L|$, and $n^* \geq 1$. Now we can define $m_G = \sum_{i=1}^{n^*} v_i(1-\frac{M_{i-1}}{1-\sigma})$, which we will see is used as a lower bound for a greedy algorithm over the light agents.

    We are now ready to describe an algorithm that sets the contract for each incentivized agent $i$ to be $t^i_j = \frac{c_i}{p^i_{-1}}$ for every $j \neq 1$. We denote the expected value of this algorithm by $ALG$. The algorithm computes $m_1$ and $m_G$. If there is a remaining heavy agent and $m_1 \geq m_G$, it attempts to incentivize the heavy agent $i^*$ corresponding to $m_1$, with payment $t^{i^*}_j = \frac{c_{i^*}}{p^{i^*}_{-1}}$ for every $j \neq 1$. If its outcome is $o_1$ it recomputes $m_1$ on the remaining agents, and continues with the algorithm on the remaining agents. If its outcome is not $o_1$, it halts. 
    
    If $m_G > m_1$ or there are no remaining heavy agents, the algorithm attempts to incentivize the light agents $i$ in decreasing order with payment $t^i_j = \frac{c_i}{p^i_{-1}}$ for every $j \neq 1$, until the remaining budget is insufficient to incentivize the next agent.

    \begin{claim}
        $ALG \geq \max\{m_1^{success}, m_G\}$.
    \end{claim}
    \begin{proof}
    We first prove that $ALG \geq m_G$. Note that when the algorithm reaches the greedy phase, every agent it attempted to incentivize previously is a heavy agent. Additionally, its realization is $o_1$, so it was not paid. Thus, if the greedy phase of the algorithm occurs, it begins with the full budget $B$ and all light agents available. Denote by $payment(k)$ the total payment to the first $k$ agents. Recall that by definition, the algorithm pays each light agent at most $\sigma$, so if $payment(k-1) \leq 1-\sigma$, the remaining budget is at least $\sigma \geq \frac{c_k}{p^k_{-1}}$, so the greedy algorithm incentivizes the $k$-th agent. Since agents with outcome $o_1$ receive no payment: 
    $$payment(k-1) \leq \sum_{\ell=1}^{k-1} \frac{c_\ell}{p^\ell_{-1}} \cdot \mathbf{1}[\text{agent } \ell \text{ realizes an outcome other than $o_1$ if incentivized}]$$ for every realization.
    By definition, the expected payment for the first $k-1$ items, given by the right-hand side of the equation, is $M_{k-1}$. Thus, from the Markov bound, the probability that the $k$'th agent is incentivized is at least $$\Pr(payment(k-1)\leq 1-\sigma) \geq 1- \frac{M_{k-1}}{1-\sigma}.$$
    
    Therefore, the expected value provided by light agent $k$ is at least $v_k(1-\frac{M_{k-1}}{1-\sigma})$, so the expected value of the greedy choice of light items is at least $m_G$.

    We now prove that $ALG \geq m_1^{success}$. Note that by definition, initially $m_1 \geq m_1^{success}$. As long as $m_1 \geq m_G$, the algorithm attempts to incentivize the agent $i^*$ with the highest expected value conditioned on it having an outcome other than $o_1$, $\frac{v_{i^*}}{p^{i^*}_{-1}}$. We split into two cases. Let $H = S\cap N_H$, where $S \sim S_{success}$, and fix a realization of $H$. If the heavy phase halts with an agent $\ell$ with outcome other than $o_1$, then all agents incentivized before $\ell$ had a realized outcome of $o_1$, and thus do not belong to $H$. Thus, every agent in $H$ was still available when $\ell$ was reached. Therefore, when the algorithm halts after incentivizing $\ell$, $m_1$ satisfies $m_1 = \frac{v_\ell}{p^{\ell}_{-1}}\geq \max_{i \in H} \frac{v_i}{p^i_{-1}}$, and the value achieved by the algorithm is $\frac{v_\ell}{p^\ell_{-1}}$ in expectation. 
    
    If instead the algorithm switches to the greedy phase, then either
$m_G>m_1$, or there are no remaining heavy agents, in which case $H$ is empty. At that point, all agents tried in the heavy phase had outcome $o_1$, so all agents in $H$ remain available, and $m_1 \geq \max_{i \in H} \frac{v_i}{p^i_{-1}}$. Thus $m_G \geq m_1 \geq \max_{i \in H} \frac{v_i}{p^i_{-1}}$, and since the realizations of the light agents are independent of $H$, the  expected value of the greedy phase conditioned on $H$ is at least $m_G$, so we again obtain that the expected value of the algorithm conditioned on $H$ is at least $\max_{i \in H} \frac{v_i}{p^i_{-1}}$. In both cases, for every realization $H$, the expected value of the algorithm conditioned on $H$ is at least $\max_{i \in H} \frac{v_i}{p^i_{-1}}$, so by taking the expectation over $H$ we get $ALG \geq m_1^{success}$.
\end{proof}
   
   Since our algorithm uses the same local contract-selection rule for every agent, it suffices to establish a constant approximation ratio between $ALG$ and $OPT$. In order to do this, we define the fractional knapsack problem over the light items by: $$\Phi (b) = \max\{\sum_{i \in N_L}v_ix_i: 0\leq x_i \leq 1, \sum_{i \in N_L}c_i x_i \leq b\}$$
    The fractional knapsack problem has a known solution: $$\forall M_{k-1} \leq x < M_{k}; \quad  \Phi(x) = \sum_{i=1}^{k-1} v_i + \frac{x-M_{k-1}}{c_k}v_k$$ Thus, it is piecewise linear and concave.

    As we did in the proof of Theorem \ref{thm:p+q=1}, we fix an optimal adaptive algorithm for maximizing the value under the budget, and let $S_{OPT}$ be the distribution of the sets of agents incentivized by it. We can use this distribution in order to provide an upper bound for its expected value $OPT$ when all agents are light.
    \begin{lemma}
    \label{lem:all_light}
        If all agents are light, $OPT \leq \Phi(1)$.
    \end{lemma}
    \begin{proof}
    
        Fix an optimal adaptive algorithm, and let $S_{OPT}$ be the distribution of the sets of agents incentivized by it. Let $x_i \in [0,1]$ be the probability agent $i$ is incentivized in the algorithm, i.e. $x_i = \Pr_{S \sim S_{OPT}}[i \in S]$. As the expected payment to each incentivized agent is at least $c_i$, the expected value is $OPT \leq \sum_i x_iv_i$. Additionally, since the expected cost must be under budget, $\sum_i x_ic_i \leq 1$. Therefore, $(x_i)$ is a feasible solution to the fractional knapsack problem over the light items with objective value at least $OPT$, while $\Phi(1)$ is the maximum over all feasible solutions, giving $OPT \leq \Phi(1)$.
    \end{proof}
    
    We can also use $\Phi(1)$ to show a lower bound for $m_G$.
    \begin{lemma}
    \label{lem:phi_gm_bound}
        $m_G \geq \frac{1- \sigma}{2}\Phi(1)$
    \end{lemma}
    \begin{proof}
        If there are no light agents, $m_G=\Phi(1)=0$, so the inequality holds trivially. We can therefore assume that the set of light agents is nonempty, so $n^* \geq 1$.
        
        Recall $m_G = \sum_{i=1}^{n^*} v_i(1-\frac{M_{i-1}}{1-\sigma}) = \frac{1}{1-\sigma}\sum_{i=1}^{n^*}v_i(1-\sigma - M_{i-1})$. We define $m'_{G}$ such that $m_G = \frac{1}{1-\sigma}m'_{G}$.

        We show a lower bound for $m'_{G}$. By definition, the summation of elements in $m'_{G}$ can be considered as the sum of areas of rectangles with a range of $[M_{i-1}, 1-\sigma]$ in the $x$-axis, and a range of $[\Phi(M_{i-1}) , \Phi(M_i)] = [\sum_{\ell=1}^{i-1} v_\ell, \sum_{\ell=1}^i v_\ell]$ in the $y$-axis. The rectangles can be seen in Figure \ref{fig:rectangles}.

        \begin{figure}[h]
  \centering
  \begin{tikzpicture}[scale=1]

    \fill[gray!25] (0,0)     rectangle (6.5,2);
    \fill[gray!25] (1,2)     rectangle (6.5,3.5);
    \fill[gray!25] (2.5,3.5) rectangle (6.5,4.8);
    \fill[gray!25] (4,4.8)   rectangle (6.5,5.8);
    \fill[gray!25] (5.5,5.8) rectangle (6.5,6.5);

    \draw (0,0)     rectangle (6.5,2);
    \draw (1,2)     rectangle (6.5,3.5);
    \draw (2.5,3.5) rectangle (6.5,4.8);
    \draw (4,4.8)   rectangle (6.5,5.8);
    \draw (5.5,5.8) rectangle (6.5,6.5);

    \draw[thick, dashed] (0,0) -- (6.5,0) -- (6.5,6.5) -- cycle;

    \draw[thick] (0,0) -- (1,2) -- (2.5,3.5) -- (4,4.8) -- (5.5,5.8) -- (6.5,6.5);

    \foreach \x/\y in {1/2, 2.5/3.5, 4/4.8, 5.5/5.8, 6.5/6.5} {
      \draw[dotted] (\x,0) -- (\x,\y);
    }

    \draw[->] (-0.3,0) -- (8.2,0) node[right] {$c$};
    \draw[->] (0,-0.3) -- (0,8)   node[above] {};

    \foreach \x/\lbl in {1/$M_1$, 2.5/$M_2$, 4/$M_3$, 5.5/$M_4$} {
      \draw (\x, 0.06) -- (\x,-0.06);
      \node[below, font=\small] at (\x,-0.15) {\lbl};
    }
    \draw (6.5, 0.06) -- (6.5,-0.06);
    \node[below, font=\small] at (6.5,-0.15) {$V$};
    \node[above right, font=\small] at (6.5,0.1) {$1{-}\sigma$};

    \draw (-0.06,6.5) -- (0.06,6.5);
    \node[left, font=\small] at (-0.1,6.5) {$\Phi(1{-}\sigma)$};

    \node[below left]  at (0,0)     {$W$};
    \node[above right] at (6.5,6.5) {$U$};

    \node[font=\scriptsize] at (3.25,1.0)  {$v_1(1{-}\sigma{-}M_0)$};
    \node[font=\scriptsize] at (3.75,2.75) {$v_2(1{-}\sigma{-}M_1)$};
    \node[font=\scriptsize] at (4.5,4.15)  {$v_3(1{-}\sigma{-}M_2)$};
    \node[font=\scriptsize] at (5.25,5.3)  {$v_4(1{-}\sigma{-}M_3)$};

    \node[font=\small] at (2.5,5.8) {$\Phi(c)$};

  \end{tikzpicture}
  \caption{A graph of the fractional knapsack solution $\Phi(c)$. The dashed triangle $UVW$ has area $\frac{1-\sigma}{2}\Phi(1-\sigma)$. The light shaded area represents $m'_G=(1-\sigma)m_G$, which fully covers the triangle.}
  \label{fig:rectangles}
\end{figure}
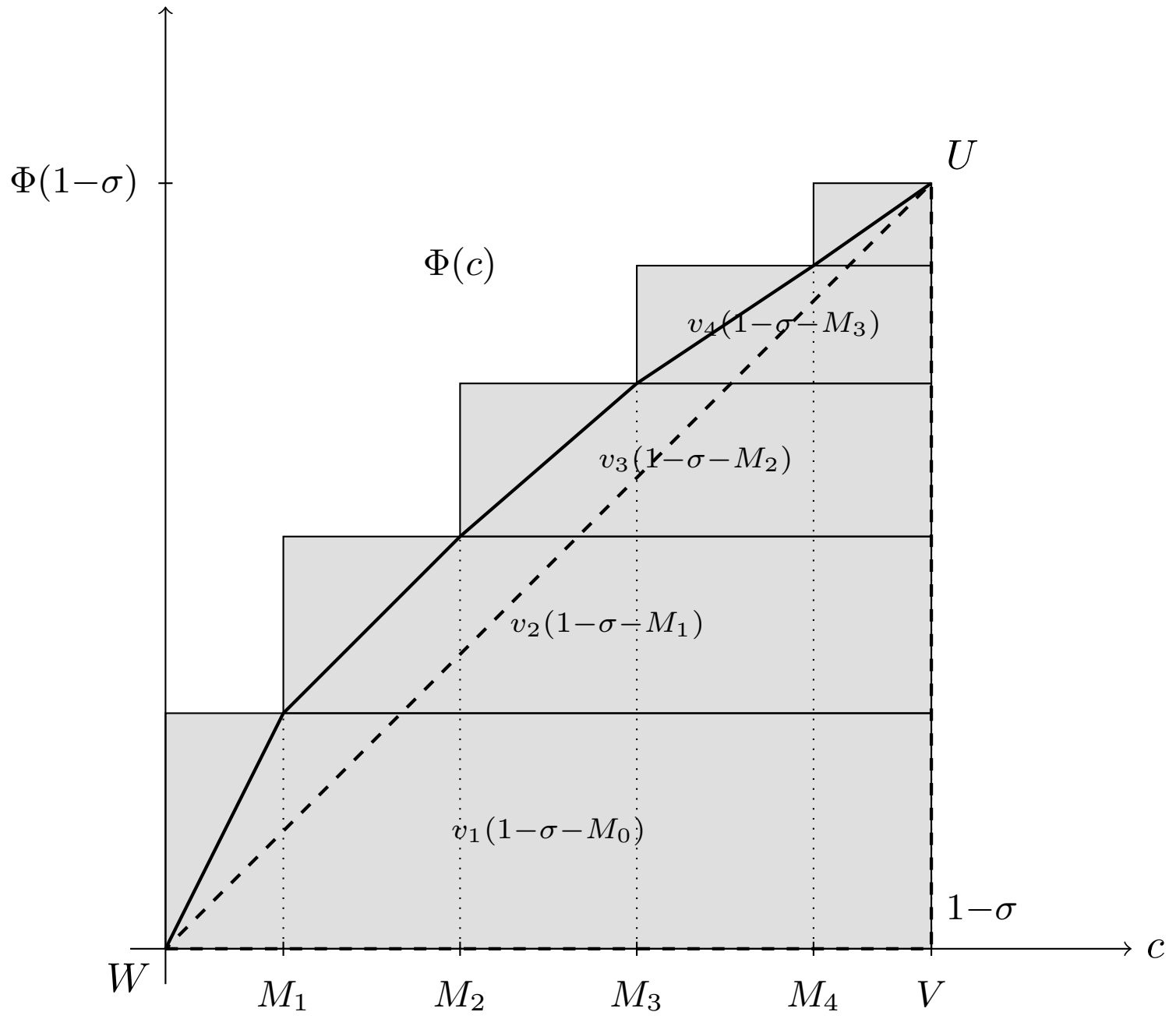
        
        Consider a triangle with vertices $W=(0,0), V=(1-\sigma, 0), U=(1-\sigma, \Phi(1-\sigma))$. For any point $(x,y)$ in this triangle, $y \leq \frac{\Phi(1-\sigma)}{1-\sigma}x$. By maximality of $n^*$, either $n^* = |N_L|$ in which case $\Phi(M_{n^*}) = \Phi(1)$ and the rectangles cover all $y \in [0, \Phi(M_{n^*})]$, or $n^* < |N_L|$, in which case $M_{n^*} > 1-\sigma$, so $\Phi(M_{n^*}) \geq \Phi(1-\sigma)$. In both cases, every point $(x,y)$ in the triangle satisfies $y \leq \Phi(1-\sigma) \leq \Phi(M_{n^*})$, so there exists $i\leq n^*$ with $\Phi(M_{i-1}) \leq y \leq \Phi(M_i)$. For this rectangle $i$, if $i=1$ then $x \geq 0=M_0$ holds trivially. Otherwise, concavity of $\Phi$ with $\Phi(0) = 0$ gives $\frac{\Phi(M_{i-1})}{M_{i-1}} \geq \frac{\Phi(1-\sigma)}{1-\sigma}$, so $x \geq \frac{y(1-\sigma)}{\Phi(1-\sigma)} \geq M_{i-1}$. In both cases $(x,y)$ is inside rectangle $i$. Thus, the rectangles cover the area of the triangle, giving $m'_{G} \geq \frac{1-\sigma}{2}\Phi(1-\sigma)$. From concavity of $\Phi$, $m_G =\frac{m'_{G}}{1-\sigma} \geq \frac{\frac{(1-\sigma)^2\Phi(1)}{2}}{1-\sigma}=\frac{1- \sigma}{2}\Phi(1)$.
    \end{proof}

    Combining Lemmas~\ref{lem:all_light} and~\ref{lem:phi_gm_bound}, we get the following stronger result when all agents are light:
    
    \begin{observation}
        If all agents are light, $ALG\geq\frac{1-\sigma}{2}OPT$.
    \end{observation}
    \begin{proof}
        From Lemma \ref{lem:all_light} $OPT \leq \Phi(1)$, and combining it with Lemma \ref{lem:phi_gm_bound} we get $OPT \leq \frac{2m_G}{1- \sigma}$. Since $ALG \geq m_G$ when all agents are light, we get $$ALG \geq \frac{1-\sigma}{2}OPT.$$
    \end{proof}
    
    However, we still need to incorporate heavy agents. First, we prove a bound for light items even when heavy agents are included.

    \begin{lemma}
    \label{lem:light_bound}
    $\EX_{S \sim S_{OPT}}(\sum_{i \in S \cap N_L} v_i) \leq \frac{2}{1-\sigma}m_G$
    \end{lemma}
    \begin{proof}
        We prove that the total expected value of incentivized light agents is at most $\frac{2}{1-\sigma}m_G$. 

        For $S \sim S_{OPT}$, let $S_L = S\cap N_L$ be the light agents in a realization. Since $S_L$ is a fractional knapsack solution over the light agents with a knapsack capacity of $\sum_{i \in S_L} c_i$ we have: 
        $$\EX_{S \sim S_{OPT}}(\sum_{i \in S_L} v_i) \leq \EX_{S \sim S_{OPT}}(\Phi(\sum_{i \in S_L}c_i))$$
        Since $\Phi$ is concave, by Jensen's inequality we get:
        $$\EX_{S \sim S_{OPT}}(\Phi(\sum_{i \in S_L}c_i)) \leq \Phi(\EX_{S \sim S_{OPT}}(\sum_{i \in S_L}c_i)).$$ Lastly, since $\EX_{S \sim S_{OPT}}(\sum_{i \in S_L}c_i) \leq 1$  and from monotonicity of $\Phi$:
        $$\Phi(\EX_{S \sim S_{OPT}}(\sum_{i \in S_L}c_i)) \leq \Phi(1)$$
        Lemma \ref{lem:phi_gm_bound} completes the proof.
    \end{proof}
    
We now only need to provide an upper bound on the expected value of heavy agents in $S_{OPT}$.
\begin{lemma}
\label{lem:heavy_bound}
        $\EX_{S \sim S_{OPT}}(\sum_{i \in S \cap N_H} v_i) \leq 2(\lceil\frac{1}{\sigma}\rceil -1)m^{success}_1$.
\end{lemma}
\begin{proof}
    We need to prove an upper bound for the expected value of the sum of all heavy agents in the optimal algorithm. We define $OPT_{hk}$ to be the expected value of an optimal adaptive algorithm over the heavy agents, that knows which agents, when incentivized, have a realized outcome of $o_1$. Clearly, $\EX_{S \sim S_{OPT}}(\sum_{i \in S \cap N_H} v_i) \leq OPT_{hk}$, as the additional information can only increase the expected value of the algorithm. Note that the agents' incentive constraints are unaffected by the algorithm's oracle access, and therefore the expected payment to each agent $i$ is still at least  $c_i$.

        Condition on a realization $S\sim S_{success}$. As the algorithm knows which
    agents have a nonzero outcome, each agent $i\in S\cap N_H$ can be viewed as
    an agent $i'$ with
    $p^{i'}_{2,1}=0$,
    $p^{i'}_{2,j}=\frac{p^i_{2,j}}{p^i_{-1}}$ for every $j\neq1$,
    $c_{i'}=\frac{c_i}{p^i_{-1}}$, and expected value
    $\frac{v_i}{p^i_{-1}}$.

    By the fractional-knapsack argument in the proof of
    Theorem~\ref{thm:p+q=1}, conditioned on $S$, the optimal adaptive algorithm has an expected value which is at
    most twice the optimal integer knapsack value in which agent $i$ has size
    $\frac{c_i}{p^i_{-1}}$ and value $\frac{v_i}{p^i_{-1}}$. Since every heavy
    agent satisfies $\frac{c_i}{p^i_{-1}}>\sigma$, such a knapsack solution
    contains at most $\lceil\frac{1}{\sigma}\rceil-1$ agents. Taking expectation over $S \sim S_{success}$, we get
    \[
        OPT_{hk}
        \leq
        2\left(\left\lceil\frac{1}{\sigma}\right\rceil-1\right)
        \EX_{S\sim S_{success}}
        \left(\max_{i\in S\cap N_H}\frac{v_i}{p^i_{-1}}\right)
        =
        2\left(\left\lceil\frac{1}{\sigma}\right\rceil-1\right)m^{success}_1.
    \]

    Thus, putting all inequalities together, $\EX_{S \sim S_{OPT}}(\sum_{i \in S \cap N_H} v_i) \leq 2(\lceil\frac{1}{\sigma}\rceil -1)m^{success}_1$
\end{proof}

We are now ready to prove a $\frac{1}{6}$-approximation to $OPT$ using local contracts.
\begin{theorem}[General case]
\label{thm:riskfree_gen}
$ALG \geq \frac{1}{6}OPT$.
\end{theorem}
\begin{proof}
Since all agents are either light or heavy, from Lemma \ref{lem:light_bound}, Lemma \ref{lem:heavy_bound}, and the fact that
$ALG \geq \max\{m_G,m^{success}_1\}$, we get
\[
ALG
\geq
\frac{
\frac{2}{1-\sigma}m_G+
2(\lceil\frac{1}{\sigma}\rceil-1)m^{success}_1
}{
\frac{2}{1-\sigma}+2\lceil\frac{1}{\sigma}\rceil-2
}
\geq
\frac{OPT}{
\frac{2}{1-\sigma}+2\lceil\frac{1}{\sigma}\rceil-2
}.
\]

Choosing $\sigma=\frac{1}{2}$ gives $ALG \geq \frac{1}{6}OPT$.
\end{proof}

Thus, although local contracts can be arbitrarily bad in general, the
risk-free structure is sufficient to obtain a polynomial-time local policy
achieving a constant fraction of the optimal adaptive value.

\subsection{Min-Pay Contract}
\label{sec:minpay}
Local contracts are not the only natural way to simplify contract design.
In this section we consider the min-pay contract, which minimizes the expected payment needed to induce effort. Note that this is an adaptive contract, as the maximal payment for an outcome is bounded by the remaining budget. In this section, we denote by $ALG$ the expected value of the algorithm which attempts to incentivize each agent using the min-pay contract. Throughout this subsection, agents are identical
and binary-action, and we normalize $B=1$.

An interesting structural property of the min-pay contract is that, given a remaining budget of
$b$, there is at most one outcome with a payment other than $0$ or $b$. To see this, note first that paying an outcome with $p_{2,j} \leq p_{1,j}$ only weakens the incentive constraint while
increasing the expected payment, so such outcomes receive no payment. For the remaining
outcomes, consider the ratio $\frac{p_{2,j}-p_{1,j}}{p_{2,j}}$, which is the incentive
provided by a unit of expected payment on outcome $o_j$. If some outcome with a higher ratio
receives a payment smaller than $b$ while an outcome with a lower ratio receives a positive
payment, then increasing the former and decreasing the latter while keeping the incentive
constraint unchanged strictly decreases the expected payment. Thus the outcomes are paid in
decreasing order of this ratio, breaking ties by order of outcomes, each receiving $b$ until the incentive constraint is satisfied,
and the last paid outcome receives the amount required for equality.

We show that the performance of the min-pay contract changes dramatically with the number of outcomes that can be used to incentivize effort, which is the number of outcomes with $p_{2,j} > p_{1,j}$. When there is only one such
outcome, using the min-pay contract is optimal. When there are two, we guarantee an approximation ratio of
$\frac{1-e^{-2}}{2}\approx 0.432$, while no approximation ratio greater than
$\frac{1}{2}$ is possible in general, and when there are three or more such outcomes, no
positive constant approximation is possible. We therefore get these approximation ratios as a function of the number of outcomes $m$:

\begin{center}
\begin{tabular}{l|l}
    case & min-pay approximation ratio\\
    \hline
    $m=2$ & $1$ \\
    $m=3$ & between $\frac{1-e^{-2}}{2}$ and $\frac{1}{2}$ \\
    $m \geq 4$ & arbitrarily close to $0$
\end{tabular}
\end{center}

\begin{claim}
\label{clm:minpay_one_positive}
    If exactly one outcome satisfies $p_{2,j} > p_{1,j}$, incentivizing agents via the min-pay contract is optimal. In
    particular, this holds for every instance with $m=2$.
\end{claim}
\begin{proof}
    Denote by $o_{j^*}$ the outcome with $p_{2,j^*} > p_{1,j^*}$. Since every other outcome
    contributes non-positively to the incentive constraint, every contract which incentivizes
    an agent satisfies $p_{2,j^*}t_{j^*} -c \geq p_{1,j^*}t_{j^*}$, so
    $t_{j^*} \geq \frac{c}{p_{2,j^*}-p_{1,j^*}}$. The min-pay contract pays exactly this amount
    to $o_{j^*}$ and nothing to the other outcomes, so its payment is at most that of any other
    incentivizing contract on every outcome. Thus, it is optimal.
\end{proof}

If there are two outcomes $j$ with $p_{2,j} > p_{1,j}$, the min-pay contract prefers the one with the higher ratio, even when paying it exhausts the budget while spreading the payment over both would leave enough for additional agents. The following
claim shows that this can cost a factor arbitrarily close to $2$.

\begin{claim}
\label{clm:minpay_lower}
    For every $\alpha>\frac{1}{2}$ there is an instance with $m=3$ outcomes and $n=3$ identical agents on which
    $ALG<\alpha\cdot OPT$.
\end{claim}
\begin{proof}
    Fix $\delta \in (0,\frac{1}{3})$. Let $r_1 = r_2 = 0$ and $r_3 = 1$, and consider three identical agents with $p_2 = (\delta, \delta, 1-2\delta)$,
    $p_1 = (1-\frac{\delta}{2}, \frac{\delta}{2}, 0)$ and
    $c = \frac{1}{2}(1-\frac{3\delta}{2})$. Then $p_{2,1} - p_{1,1} = -(1-\frac{3\delta}{2})$,
    $p_{2,2} - p_{1,2} = \frac{\delta}{2}$ and $p_{2,3} - p_{1,3} = 1-2\delta$, so every
    incentivized agent has an expected value of $v = 1-2\delta$ while an unincentivized agent
    has an expected value of $0$, and the incentive constraint under a contract $t$ is
    $\delta t_1 + \delta t_2 + (1-2\delta)t_3-c \geq (1- \frac{\delta}{2}) t_1 + \frac{\delta}{2} t_2$.

    \begin{table}[h]
  \centering
  \begin{tabular}{l|ccc|c}
     & $o_1$ & $o_2$ & $o_3$ & $c$ \\
    \hline
    Action $1$ & $1-\frac{\delta}{2}$ & $\frac{\delta}{2}$ & $0$ & $0$\\
    Action $2$ & $\delta$ & $\delta$ & $1-2\delta$ & $\frac{1}{2}(1-\frac{3\delta}{2})$ \\
    \hline
    $r_j$ & $0$ & $0$ & $1$ &\\
  \end{tabular}
\end{table}

    Since $p_{2,1} < p_{1,1}$, an optimal contract sets $t_1 = 0$. Ex-post feasibility with a
    remaining budget of $b$ requires $t_j \leq b$ for every $j$, so an agent can be
    incentivized with a remaining budget of $b$ if and only if $b(1-\frac{3\delta}{2}) \geq c = \frac{1}{2}(1- \frac{3\delta}{2})$,
    that is, if and only if $b \geq \frac{1}{2}$.

    The min-pay contract pays the outcomes in decreasing order of the ratio
    $\frac{p_{2,j}-p_{1,j}}{p_{2,j}}$, which is $1$ for $o_3$ and $\frac{1}{2}$ for $o_2$, so it
    pays $o_3$ first and pays $o_2$ only if the remaining budget prevents $o_3$ alone from satisfying
    the incentive constraint. Paying only $o_3$ requires
    $t_3 = \frac{c}{1-2\delta} = \frac{1-\frac{3\delta}{2}}{2(1-2\delta)}$, which satisfies
    $\frac{1}{2} < t_3 < 1$ for every $\delta > 0$. Thus with the full budget the min-pay
    contract pays $t_3$ on $o_3$ and nothing elsewhere, and once $o_3$ is realized the remaining
    budget is smaller than $\frac12$ and no later agent can be incentivized, while $o_1$ and
    $o_2$ receive no payment and leave the budget unchanged. The remaining budget is therefore
    always either $1$, in which case the same contract is offered again, or smaller than
    $\frac12$, in which case no agent can be incentivized, so the min-pay contract never pays
    $o_2$. Since $o_3$ is realized with probability $1-2\delta$,
    $ALG = (1-2\delta)(1 + 2\delta + 4\delta^2)$.

    Consider instead the algorithm that incentivizes every agent with $t_2 = t_3 = \frac12$,
    which satisfies the incentive constraint with equality since
    $\frac{1}{2}(1-\frac{3\delta}{2}) = c$. It pays $\frac12$ whenever $o_2$ or $o_3$ is
    realized, which occurs with probability $1-\delta$ and leaves a remaining budget of exactly
    $\frac12$, and pays nothing otherwise. Thus the first two agents are always incentivized,
    and the third is incentivized unless both earlier agents were paid, so
    $OPT \geq (1-2\delta)(2 + (1-(1-\delta)^2)) = (1-2\delta)(2 + 2\delta - \delta^2)$.

    Therefore $\frac{ALG}{OPT} \leq \frac{1+2\delta+4\delta^2}{2+2\delta-\delta^2}$, which
tends to $\frac{1}{2}$ as $\delta \to 0$.
\end{proof}

Note that for $m=3$ the differences $p_{2,j}-p_{1,j}$ sum to $0$, so at most two of them
are positive.

We do not have a tight bound for $m=3$. However, we can guarantee an approximation ratio of $\frac{1-e^{-2}}{2} \approx 0.432$. To prove this lower bound we first need the following probabilistic lemma, which bounds the
number of agents required for a given outcome to be realized a fixed number of times.

\begin{lemma}
\label{lem:stopping_time}
    Let $K \geq 2$, and let $T$ be the number of independent trials of success probability $q$
    required for $K$ successes. Then for every $k \geq 0$,
    $\EX[\min\{k,T\}] \geq \frac{1}{2}\min\{k,\frac{K+1}{q}\}$.
\end{lemma}
\begin{proof}
    Write $k = Ka+r$ for integers $a \geq 0$ and $0 \leq r < K$, and let $G_1, \dots, G_K$ be
    the numbers of trials between consecutive successes, which are independent and geometric
    with parameter $q$ and sum to $T$. Truncating $r$ of them at $a+1$ and the remaining $K-r$
    at $a$, the truncation levels sum to $k$, so their sum is at most both $k$ and $T$. Since
    $\EX[\min\{a,G_i\}] = \frac{1-(1-q)^a}{q}$,
    $$\EX[\min\{k,T\}] \geq (K-r)\frac{1-(1-q)^a}{q} + r\frac{1-(1-q)^{a+1}}{q} =
    \frac{K(1-(1-q)^a)}{q} + r(1-q)^a.$$
    Multiplying by $q$, and since $kq = Kaq+rq$, it suffices to prove
    $K(1-(1-q)^a) + rq(1-q)^a \geq \frac{1}{2}\min\{kq, K+1\}$.

    Suppose first that $kq \leq K+1$ and $(1-q)^a \geq \frac12$. Then
    $rq(1-q)^a \geq \frac{rq}{2}$, and $Kaq \leq kq \leq K+1$ gives $aq \leq \frac{K+1}{K} \leq
    \frac32$, so $1-(1-q)^a \geq 1-e^{-aq} \geq \frac{aq}{2}$. Adding the two inequalities
    proves the required bound.

    Otherwise, we show that the left hand side exceeds the required bound by at least
    $\frac{K-1}{2} + (1-q)^a(1-Kaq)$. If $kq \leq K+1$ and $(1-q)^a < \frac12$, then
    $rq \leq K+1-Kaq$ and $(1-q)^a - \frac12 < 0$, so
    $$K(1-(1-q)^a)+rq(1-q)^a-\frac{kq}{2} \geq
    K\left(1-(1-q)^a-\frac{aq}{2}\right) + (K+1-Kaq)\left((1-q)^a-\frac12\right),$$
    which equals $\frac{K-1}{2}+(1-q)^a(1-Kaq)$. If instead $kq > K+1$, then
    $rq \geq K+1-Kaq$ and $(1-q)^a > 0$, so
    $$K(1-(1-q)^a)+rq(1-q)^a-\frac{K+1}{2} \geq
    K(1-(1-q)^a)+(K+1-Kaq)(1-q)^a-\frac{K+1}{2} = \frac{K-1}{2}+(1-q)^a(1-Kaq).$$
    If $Kaq \leq 1$ this is non-negative, so assume $Kaq > 1$. Then $1-Kaq<0$ and
    $(1-q)^a \leq e^{-aq}$, so it is at least $\frac{K-1}{2}+e^{-aq}(1-Kaq)$. The function
    $\frac{K-1}{2}+e^{-s}(1-Ks)$ has derivative $e^{-s}(Ks-K-1)$ and is therefore minimized at
    $s = 1+\frac1K$, where it equals $\frac{K-1}{2}-Ke^{-1-\frac1K}$. This is positive for
    $K=2$ and $K=3$ by direct computation, and for $K \geq 4$ we have
    $2Ke^{-1-\frac1K} < \frac{2K}{e} \leq K-1$.
\end{proof}

We are now ready to prove a lower bound on the approximation ratio of the min-pay contract.
\begin{claim}
\label{clm:minpay_upper}
    For $m=3$, $ALG\geq \frac{1-e^{-2}}{2}OPT$.
\end{claim}
\begin{proof}
    If only one outcome satisfies $p_{2,j}>p_{1,j}$ the claim follows from Claim
    \ref{clm:minpay_one_positive}, so assume there are exactly two, denoted $o_2,o_3$ and
    ordered so that $\frac{p_{2,2}-p_{1,2}}{p_{2,2}} \geq \frac{p_{2,3}-p_{1,3}}{p_{2,3}}$.
    Since the three differences sum to $0$, measuring all payments and budgets in units of
    $\frac{c}{(p_{2,2}-p_{1,2})+(p_{2,3}-p_{1,3})}$ and writing
    $\alpha = \frac{p_{2,2}-p_{1,2}}{(p_{2,2}-p_{1,2})+(p_{2,3}-p_{1,3})}$, the incentive
    constraint becomes $\alpha t_2 + (1-\alpha)t_3 - t_1 \geq 1$. As the coefficient of $t_1$ is
    negative, the optimal
    contract sets $t_1 = 0$. The coefficients of $t_2$ and $t_3$ sum to $1$, so under a
    remaining budget of $b$ the largest value the left hand side attains is $b$, and an agent
    can be incentivized if and only if $b \geq 1$.

    Let $OPT_k(b)$ and $ALG_k(b)$ denote the expected number of agents incentivized by an
    optimal algorithm and by the min-pay contract respectively, when $k$ agents remain and the
    remaining budget is $b$. Since a larger remaining budget allows every contract that a
    smaller one allows, both are non-decreasing in $b$, and since every incentivized agent has
    the same expected value, it suffices to compare them. The ordering of the outcomes gives
    $\frac{p_{2,2}}{\alpha} \leq \frac{p_{2,3}}{1-\alpha}$, so every contract which incentivizes
    an agent has an expected payment of $p_{2,2}t_2+p_{2,3}t_3 \geq
    \frac{p_{2,2}}{\alpha}(\alpha t_2 + (1-\alpha)t_3) \geq \frac{p_{2,2}}{\alpha}$. As the
    total payment is at most $b$ and at most $k$ agents remain,
    $OPT_k(b) \leq \min\{k, \frac{\alpha b}{p_{2,2}}\}$.

    Suppose first that $1 \leq b < \frac{1}{\alpha}$. The min-pay contract then pays $t_2 = b$
    and $t_3 = \frac{1-\alpha b}{1-\alpha}$, which is at most $b$ since $b \geq 1$, so no budget
    remains after $o_2$, while after $o_3$ the remaining budget is $\frac{b-1}{1-\alpha}$, which
    is smaller than $\frac{1}{\alpha}$ since $\alpha b < 1$. We prove by induction on $k$ that
    $ALG_k(b) \geq \frac{1}{2}OPT_k(b)$ for every such $b$. By symmetry, we may assume that the optimal algorithm incentivizes the first agent
whenever it incentivizes any agent. If it incentivizes no agent, the claim is trivial. Consider the
    contract $t$ used by an optimal algorithm when there are $k$ remaining agents, and a remaining budget of $b$. Reducing payments preserves ex-post feasibility
    and only increases the remaining budgets, so we may assume $t$ satisfies its incentive
    constraint with equality, that is $\alpha(b-t_2)+(1-\alpha)(b-t_3) = b-1$. Therefore
    $b-t_3 \leq \frac{b-1}{1-\alpha}$ and $b-t_2 \leq b < \frac{1}{\alpha}$, and in particular
    $p_{2,2}OPT_{k-1}(b-t_2) \leq \alpha(b-t_2) < 1$. Since $o_1$ receives no payment,
   $$\frac{1}{2}OPT_k(b)
\leq
1+\frac{p_{2,1}}{2}OPT_{k-1}(b)
+\frac{p_{2,3}}{2}OPT_{k-1}(\frac{b-1}{1-\alpha}),$$
which by the induction hypothesis is at most
$$1+p_{2,1}ALG_{k-1}(b)
+p_{2,3}ALG_{k-1}(\frac{b-1}{1-\alpha})
=ALG_k(b),$$ where the
    induction hypothesis is applied to $\frac{b-1}{1-\alpha}$ only when it is at least $1$, as
    otherwise both sides vanish.

    Now suppose $b \geq \frac{1}{\alpha}$ and let $K = \lfloor \alpha b \rfloor \geq 1$. The
    min-pay contract then pays $t_2 = \frac{1}{\alpha}$ and $t_1=t_3=0$. As long as $o_2$ has
    been realized fewer than $K$ times the remaining budget is at least
    $b - \frac{K-1}{\alpha} \geq \frac{1}{\alpha} \geq 1$, so the contract remains feasible
    until $o_2$ has been realized $K$ times. Letting $T$ be the number of agents required for
    this, $ALG_k(b) \geq \EX[\min\{k,T\}]$. If $K \geq 2$, then since $\alpha b < K+1$, Lemma
    \ref{lem:stopping_time} applied with success probability $p_{2,2}$ gives
    $$ALG_k(b) \geq \EX[\min\{k,T\}]
\geq \frac{1}{2}\min\{k,\frac{K+1}{p_{2,2}}\}
\geq \frac{1}{2}OPT_k(b).$$

    If $K=1$, then $\alpha b < 2$ and $T$ is geometric, so
    $ALG_k(b) \geq \frac{1-(1-p_{2,2})^k}{p_{2,2}}$ while
    $OPT_k(b) \leq \min\{k,\frac{2}{p_{2,2}}\}$. Using $(1-p_{2,2})^k \leq e^{-kp_{2,2}}$,
    $$\frac{ALG_k(b)}{OPT_k(b)} \geq
\frac{1-(1-p_{2,2})^k}{\min\{kp_{2,2},2\}}
\geq
\frac{1-e^{-kp_{2,2}}}{\min\{kp_{2,2},2\}}
\geq
\frac{1-e^{-2}}{2},$$
where the last inequality follows since
$\frac{1-e^{-kp_{2,2}}}{kp_{2,2}}$ is decreasing for
$kp_{2,2}\leq 2$, while
$\frac{1-e^{-kp_{2,2}}}{2}$ is increasing for
$kp_{2,2}\geq2$, so the minimum is attained at $kp_{2,2}=2$.
As $\frac{1-e^{-2}}{2}<\frac12$, this bound holds in every case.
\end{proof}

For the case of $m\geq 4$, there may be three or more outcomes $j$ with positive $p_{2,j} - p_{1,j}$. We show that in this case the approximation ratio of the min-pay contract can be arbitrarily close to $0$.

\begin{claim}
\label{clm:minpay_unbounded}
    For every $\alpha>0$ there is an instance with $m=4$ outcomes and identical agents on which
$ALG < \alpha \cdot OPT$.
\end{claim}
\begin{proof}
\begin{table}[h]
  \centering
  \begin{tabular}{l|cccc|c}
     & $o_1$ & $o_2$ & $o_3$ & $o_4$ & $c$ \\
    \hline
    Action $1$ & $\frac{1}{2}+\delta^2$ & $\delta-\delta^2$ & $\frac{1}{2}-\delta$ & $0$ & $0$\\
    Action $2$ & $\frac{1}{2}-2\delta$ & $\delta$ & $\frac{1}{2}$ & $\delta$ & $\delta+\delta^2$ \\
    \hline
    $r_j$ & $0$ & $0$ & $0$ & $1$ &\\
  \end{tabular}
\end{table}
    Fix $\delta \in (0,\frac{1}{10})$, let $n \geq \frac{1}{\delta}$, let $r_1=r_2=r_3=0$ and
    $r_4=1$, and let the agents be identical with
    $p_2 = (\frac{1}{2}-2\delta, \delta, \frac{1}{2}, \delta)$,
    $p_1 = (\frac{1}{2}+\delta^2, \delta-\delta^2, \frac{1}{2}-\delta, 0)$ and
    $c = \delta+\delta^2$. The differences $p_{2,j}-p_{1,j}$ are $-(2\delta+\delta^2)$,
    $\delta^2$, $\delta$ and $\delta$. An incentivized agent realizes $o_4$ with probability
    $\delta$ and therefore has an expected value of $\delta$, while an unincentivized agent
    never realizes $o_4$ and has an expected value of $0$.

    The min-pay contract pays nothing on $o_1$, and pays the remaining outcomes in decreasing
    order of $\frac{p_{2,j}-p_{1,j}}{p_{2,j}}$, which is $1$ for $o_4$, $2\delta$ for $o_3$ and
    $\delta$ for $o_2$. Since $\delta b < \delta + \delta^2$ for every $b \leq 1$, paying $o_4$
    alone never satisfies the incentive constraint, so with a remaining budget of $b$ the
    min-pay contract pays $t_4 = b$ and continues with $o_3$, which requires
    $t_3 = 1+\delta-b$. This is at most $b$ exactly when $b \geq \frac{1+\delta}{2}$, in which
    case nothing is paid on $o_2$.

    We bound the number of agents the min-pay contract incentivizes. Whenever $o_4$ is realized
    the entire remaining budget is paid, so no additional agent can be incentivized. Whenever $o_3$
    is realized while $b \geq \frac{1+\delta}{2}$, the remaining budget becomes
    $b - t_3 = 2b-1-\delta$, so the quantity $1+\delta-b$ doubles, and as it equals $\delta$
    with the full budget it equals $2^k\delta$ after $k$ such realizations. The contract
    therefore remains in this regime for at most $\log_2\frac{1}{\delta}$ realizations of $o_3$,
    after which one further realization of $o_3$ or $o_4$ leaves no budget, as both $o_3$ and
    $o_4$ are then paid the entire remaining budget. Thus at most
    $\log_2\frac{1}{\delta}+2$ incentivized agents realize $o_3$ or $o_4$. Each incentivized
    agent realizes one of them with probability $\frac{1}{2}+\delta$, independently of the
    others, so at most $2(\log_2\frac{1}{\delta}+2)$ agents are incentivized in expectation and
    $ALG \leq 2\delta(\log_2\frac{1}{\delta}+2)$.

    Consider instead the algorithm that incentivizes every agent with $t_2 = t_4 = 1$ and
    $t_1 = t_3 = 0$, which satisfies the incentive constraint with equality since
    $\delta \cdot 1 + \delta \cdot 1 - (\delta + \delta^2) = \delta-\delta^2 = 1\cdot(\delta-\delta^2)-0$. It is feasible with the full budget, and pays only when $o_2$ or
    $o_4$ is realized, which occurs with probability $2\delta$. Thus every agent is incentivized
    until the first payment, and since $n \geq \frac{1}{\delta}$,
    $OPT \geq \delta\frac{1-(1-2\delta)^n}{2\delta} \geq \frac{1-e^{-2}}{2}$.

    Therefore $\frac{ALG}{OPT}
\leq
\frac{4\delta(\log_2\frac1\delta+2)}{1-e^{-2}},$ which converges to $0$ as $\delta\to0$.
\end{proof}

By adding outcomes which are realized with probability $0$ under both actions, the same
instance shows that the approximation ratio of using the min-pay contract can be arbitrarily close to $0$ for every
$m \geq 4$.

The min-pay contract therefore provides a simple adaptive rule with sharply
different guarantees depending on the number of outcomes. In particular, its
failure for four or more outcomes does not imply that the optimal adaptive
policy is computationally difficult.

\subsection{Computational Tractability for Identical Binary-Action Agents}
\label{sec:identical_fptas}
Preceding results show that several natural simple contracts can perform
arbitrarily poorly even for identical agents. This might lead one to the conclusion that identical agents do not significantly simplify the problem. We show that this is not the case, and in fact, for identical agents with binary-action we can compute an arbitrarily good approximation to the
optimal adaptive policy.

While using simple contracts failed, we can still achieve an FPTAS for identical agents with binary-action using a
dynamic programming approach with rounding.

Recall that $v$ denotes the expected additional value from an agent exerting effort. We normalize $B=1$, and for simplicity we assume $c>0$, $v>0$, and that there is sufficient budget to
incentivize at least one agent; the remaining cases are trivial. Since the agents are identical,
a realization affects an adaptive algorithm only through the remaining budget.

We may represent any adaptive algorithm $A$ over $n$ agents by the contract it offers the first
agent, together with its continuations $A_1, \dots, A_m$, where $A_j$ is the algorithm over the
remaining $n-1$ agents it runs after outcome $o_j$. This is sufficient since all agents are
identical and the value function is additive. We denote by $b(A)$ the minimum budget needed to
implement the action choices and continuations of $A$, and by $v(A)$ its expected additional value from incentivized agents,
so that $v(A) = v + \sum_j p_{2,j}v(A_j)$ whenever $A$ incentivizes its first agent. 

\begin{lemma}
\label{lem:budget_recursion}
    Let $A$ incentivize its first agent and have continuations $A_1, \dots, A_m$, denote
    $b_j = b(A_j)$, and let $b'$ solve
    $\sum_{j : p_{2,j} > p_{1,j}}(p_{2,j}-p_{1,j})(b'-b_j) = c$. Then
    $b(A) = \max\{\max_j b_j, b'\}$.
\end{lemma}
\begin{proof}
    Suppose $A$ is feasible with a budget $b$, using the contract $t$ for the first agent.
    After outcome $o_j$ the remaining budget is $b-t_j$, and $A_j$ must be feasible with it, so
    $t_j \leq b-b_j$, and in particular $b \geq b_j$ since $t_j \geq 0$. In the incentive
    constraint $\sum_jp_{2,j}t_j -c \geq \sum_j p_{1,j}t_j$, the terms of the outcomes with
    $p_{2,j} \leq p_{1,j}$ are non-positive, and in the remaining terms the coefficient is
    positive, so we may substitute $t_j \leq b-b_j$ and obtain
    $c \leq \sum_{j : p_{2,j} > p_{1,j}}(p_{2,j}-p_{1,j})(b-b_j)$. The right-hand side is
    increasing in $b$ and equals $c$ at $b'$, so $b \geq b'$.

    For the other direction, let $b = \max\{\max_j b_j, b'\}$, and set $t_j = b-b_j$ for the
    outcomes with $p_{2,j} > p_{1,j}$ and $t_j = 0$ for the rest. Then $0 \leq t_j \leq b$, so
    the contract is ex-post feasible, and
    $\sum_j(p_{2,j}-p_{1,j})t_j = \sum_{j : p_{2,j} > p_{1,j}}(p_{2,j}-p_{1,j})(b-b_j) \geq c$
    since $b \geq b'$, so the first agent is incentivized. The remaining budget is $b_j$ after
    an outcome with $p_{2,j} > p_{1,j}$ and $b \geq b_j$ after every other outcome, so $A_j$ is
    feasible in both cases.
\end{proof}

Note that $b'$, and therefore also $\max\{\max_j b_j, b'\}$, is increasing in every $b_j$.

Our algorithm computes, for every number of agents $k$ and every expected additional value $x$
in a set $X$ defined below, an entry $b_k(x)$, which is a budget with which some algorithm over
$k$ agents of expected additional value at least $x$ is ex-post feasible, and is $\infty$ if no
such algorithm is found. It does so by induction on $k$, rounding the attainable values to a
grid instead of enumerating them all.

Directly using Lemma \ref{lem:budget_recursion} to obtain an entry from the previous level
requires enumerating a continuation for each of the $m$ outcomes, and the number of such choices
is exponential in $m$. To avoid this, note that the lemma gives the required budget as the
larger of $\max_j b_j$ and
$$b' = \frac{c+\sum_{j : p_{2,j}>p_{1,j}}(p_{2,j}-p_{1,j})b_j}
{\sum_{j : p_{2,j}>p_{1,j}}(p_{2,j}-p_{1,j})},$$
where the denominator is positive as otherwise no agent can be incentivized. We therefore fix
the value of $\max_j b_j$ in advance and allow only continuations whose budget is not higher.
The remaining two quantities, $b'$ and the expected value, are then both sums over the outcomes,
so the continuations can be chosen one outcome at a time, adding the contribution of each
outcome to both sums before proceeding to the next.

Fix $\sigma \in (0,1)$, and let $X = \{\frac{q\sigma v}{n(m+1)} : q\in\mathbb{Z}_{\geq0},\ q\leq
\frac{n^2(m+1)}{\sigma}\}$ be the set of values the algorithm considers, so every
$0 \leq x \leq nv$ is rounded down to the largest element of $X$ not exceeding it. The
algorithm computes the entries $b_k(x)$ for every $0 \leq k \leq n$ and $x \in X$ as follows. It
sets $b_0(0)=0$ and $b_0(x)=\infty$ for every other $x \in X$.

For $k \geq 1$, the algorithm considers every finite value $b_{k-1}$ takes on the previous
level, and restricts the continuations to the values $x_j \in X$ with $b_{k-1}(x_j)$ at most
that value. It then iterates over the outcomes one at a time, maintaining for every $x \in X$
the minimum value of $\sum_{\ell \leq j,\, p_{2,\ell}>p_{1,\ell}}
(p_{2,\ell}-p_{1,\ell})b_{k-1}(x_\ell)$ over the choices of $x_1, \dots, x_j$ which reach $x$,
which we denote by $S$. Initially only $x=0$ is reachable, with $S=0$, and iterating over $o_j$
takes a reachable $x$ and a value $x_j$ to the value $x+p_{2,j}x_j$ rounded down to $X$, with
$S$ increased by $(p_{2,j}-p_{1,j})b_{k-1}(x_j)$ if $p_{2,j}>p_{1,j}$ and unchanged otherwise.
After all $m$ outcomes have been considered, a reachable $x$ yields a sequence of contracts requiring a
budget which is the larger of the value considered and
$\frac{c+S}{\sum_{j : p_{2,j}>p_{1,j}}(p_{2,j}-p_{1,j})}$, and the algorithm assigns this budget
to the value $v+x$ rounded down to the nearest value in $X$. It then sets $b_k(x)$ to be the minimum of $b_{k-1}(x)$
and of all budgets assigned to $x$ in this way, also maintaining the choices attaining this minimum so they can be recovered. Since values are only
rounded down, the expected additional value of the algorithm recovered from an entry is at least the corresponding value. Finally, it returns the algorithm attaining
$\max\{x \in X : b_n(x) \leq 1\}$, whose expected additional value we denote by $ALG$.

\begin{lemma}
\label{lem:dp_correct}
    For every algorithm $A$ over $k$ agents there is $x \in X$ such that
    $x \geq v(A) - \frac{k\sigma v}{n}$ and $b_k(x) \leq b(A)$.
\end{lemma}
\begin{proof}
    By induction on $k$. For $k=0$ we have $v(A)=0$ and $b(A)=0$, so $x=0$ satisfies both
    requirements. For $k \geq 1$, if $A$ does not incentivize its first agent then it is also an
    algorithm over $k-1$ agents, and the claim follows from the induction hypothesis and
    $b_k(x) \leq b_{k-1}(x)$. Otherwise, each continuation $A_j$ is an algorithm over $k-1$
    agents, so by the induction hypothesis there are $x_j \in X$ with
    $x_j \geq v(A_j) - \frac{(k-1)\sigma v}{n}$ and $b_{k-1}(x_j) \leq b(A_j)$, and in
    particular $b_{k-1}(x_j)$ is finite.

    The algorithm considers the value $\max_j b_{k-1}(x_j)$, for which every $x_j$ is an allowed
    continuation. We show by induction on $j$ that after processing $o_1, \dots, o_j$ the
    algorithm retains the state obtained from $x_1, \dots, x_j$ under the successive roundings,
    with $S$ at most
    $\sum_{\ell \leq j,\, p_{2,\ell}>p_{1,\ell}}(p_{2,\ell}-p_{1,\ell})b_{k-1}(x_\ell)$. This
    holds before any outcome is processed, and if it holds after $o_1, \dots, o_{j-1}$ then
    extending that state by $x_j$ produces the required state with the required sum, which is
    retained as only the smallest such sum is kept.

    Let $x$ be the value the algorithm assigns to the resulting state after all $m$ outcomes
    have been processed. Each of the $m$ roundings during the processing of the outcomes, and
    the final rounding after adding $v$, loses at most $\frac{\sigma v}{n(m+1)}$, so since
    $\sum_j p_{2,j}=1$,
    $$x \geq v + \sum_j p_{2,j}x_j - \frac{\sigma v}{n} \geq v + \sum_j p_{2,j}v(A_j) -
    \frac{(k-1)\sigma v}{n} - \frac{\sigma v}{n} = v(A) - \frac{k\sigma v}{n}.$$
    The budget assigned to this state is the larger of $\max_j b_{k-1}(x_j)$ and
    $\frac{c+S}{\sum_{j : p_{2,j}>p_{1,j}}(p_{2,j}-p_{1,j})}$, so by Lemma
    \ref{lem:budget_recursion} and its monotonicity it is at most $b(A)$, giving
    $b_k(x) \leq b(A)$.
\end{proof}

\begin{theorem}
\label{thm:identical_fptas}
    For identical agents with binary-action, $ALG \geq (1-\sigma)OPT$, and the algorithm runs in time polynomial in $n$, $m$, and $\frac{1}{\sigma}$.
\end{theorem}
\begin{proof}
    Fix an optimal adaptive algorithm $A$ over the $n$ agents, so that $v(A)=OPT$ and
    $b(A) \leq 1$. By Lemma \ref{lem:dp_correct} there is $x \in X$ with $x \geq OPT - \sigma v$
    and $b_n(x) \leq 1$, so the algorithm returns a value of at least $x$, giving
    $ALG \geq OPT - \sigma v$. Since there is sufficient budget to incentivize at least one
    agent, $OPT \geq v$, and therefore $ALG \geq (1-\sigma)OPT$.

    For the running time, $v(A) \leq nv$ for every algorithm, so
    $|X| = O(\frac{n^2m}{\sigma})$. Computing the entries of level $k$ from those of level $k-1$
    considers at most $|X|$ values on the previous level, and for each of them processes $m$
    outcomes, where each outcome is processed by considering every state and every allowed
    continuation, which takes $O(|X|^2)$ time. With $n$ levels, the total number of arithmetic
    operations is $O(nm|X|^3)$.
\end{proof}

Thus, for identical agents with binary-action, simple contract rules may fail even though the
optimal adaptive policy admits an FPTAS.

\section{Beyond Additive Valuations}
\label{sec:beyond_additive}
In this section we move beyond additive valuations. In order to isolate the impact of nonadditivity from the other difficulties of the model, we limit our focus to binary outcomes. Initially, we also limit our focus to risk-free agents. Recall that risk-free agents are binary-action agents that deterministically lead to outcome $o_1$ when shirking.

In order to simplify notation, the principal's value can be written as $f(S)$ where $f:2^{[n]} \rightarrow \mathbb{R}_{\geq 0}$ and $S \subseteq [n]$ is the set of agents realizing the successful outcome $o_2$. We assume value-oracle access to $f$. For each agent $i$, let $p_i$ be its probability of success when exerting effort. For the AND objective, if some agent has $p_i=0$, the optimal value is $0$. For the other objectives, such agents can be discarded. We therefore assume $p_i>0$ for the remaining agents. Zero-cost agents can be incentivized with the zero contract and should always be included,
so we assume $c_i>0$ for the remaining agents.

Incentivizing agent $i$ requires a payment of $c_i/p_i$, made only upon success. We denote this payment by $t_i=c_i/p_i$.

We first consider the OR and AND objectives. As a secondary comparison, we also characterize optimal policies for maximizing the principal's utility, defined as her expected value minus her expected total payment.

\subsection{OR and AND Valuations}
\label{sec:andor}
In the OR valuation, formally, $f(S)=0$ if $S=\emptyset$ and $f(S)=1$
otherwise. Since a single success suffices,
an optimal policy for maximizing value
approaches all agents with $t_i\leq B$,
in any order, until the first success.
For the principal-utility objective, an
optimal policy instead approaches agents
with $t_i\leq\min\{B,1\}$ in increasing
order of $t_i$, again stopping after the
first success.

A non-adaptive policy fixes its agent-contract
choices independently of realized outcomes.
The price of adaptivity, defined as the ratio
between the optimal adaptive and non-adaptive
values, can be $n$ for the OR objective. Consider $B=1$ and $n$ symmetric agents with $c_i=p_i=\eps$. Since a successful agent must be paid at least $1$, only a single agent can be incentivized by a non-adaptive contract, for a value of $\eps$. However, an adaptive policy may attempt to incentivize all agents, for an expected value of $1-(1-\eps)^n \approx n\eps$.
This is tight since it is always possible to incentivize the best feasible agent and make no payment to every other agent.

We now consider the AND function, i.e.,
$f(S)=1$ iff $S=[n]$. Since every agent
must succeed, obtaining positive expected
value requires incentivizing all agents
along the all-success path. This is feasible
if and only if $\sum_i t_i\leq B$.
In that case, the optimal expected value is
$\prod_i p_i$, independently of the order
in which agents are approached. Otherwise,
the optimal value is $0$.

For the principal's utility objective, if
$\sum_i t_i>B$, the principal cannot incentivize all agents and the optimal utility is $0$.
Otherwise, all agents can be incentivized, and the probability that they all succeed is
independent of their ordering. Thus, among policies that incentivize every agent,
maximizing the principal's utility is equivalent to
minimizing expected payment. The expected payment to agent $i$ is
$(p_1\dots p_{i-1})p_i\frac{c_i}{p_i}
= p_1\dots p_{i-1}c_i$.

We must minimize $\sum_{i \leq n} c_i\prod_{j<i}p_j$ to minimize the expected payment. To do so, consider two adjacent agent indices in the ordering, $a$ and $b=a+1$. If we switch their order, the expected payment changes from  $(\prod_{i<a}p_i)(c_a+p_ac_b)$ to $(\prod_{i<a}p_i)(c_b+p_bc_a)$. Note that $c_a+p_ac_b > c_b+p_bc_a$ iff
$\frac{c_a}{1-p_a}>\frac{c_b}{1-p_b}$. Therefore, the expected payment is
minimized by ordering the agents in nondecreasing order of
$\frac{c_i}{1-p_i}$. For agents with $p_i=1$, we use the convention
$\frac{c_i}{1-p_i}=+\infty$.
The optimal policy follows this order if the resulting
expected utility is positive, and otherwise offers no
paid contracts.
\subsection{Monotone Submodular Valuations}
We next consider monotone submodular valuations. In the
risk-free binary-action setting, we show the problem is directly related to
submodular maximization under a knapsack constraint.

We also consider a stronger \emph{full-knowledge} benchmark, denoted by $OPT_{\mathrm{FK}}$, in which the principal observes in advance which agents would succeed if incentivized to exert effort. Clearly, $OPT\leq OPT_{\mathrm{FK}}$.

Let $f$ be a monotone submodular function. Note that when $p_i=1$ for every $i$, our problem reduces to the classical problem of maximizing a monotone submodular function subject to a knapsack constraint. A $1 - \frac{1}{e}$ approximation is optimal for this problem (an algorithm from \cite{Svir04}, tightness from \cite{Feige98}), so we cannot hope for a better approximation ratio in general.

It is known that under a cardinality constraint, a greedy algorithm is a $(1-\frac{1}{e})$-approximation for maximizing a monotone submodular function. Thus, as a warm-up, we consider the corresponding version of our problem, and arrive at the following interesting observation:
\begin{observation}
    For a monotone submodular $f$, if $\frac{c_i}{p_i}=1$ for every $i \in [n]$, a greedy algorithm is a $(1-\frac{1}{e})$-approximation for the full knowledge benchmark.
\end{observation}
\begin{proof}
    Let $T \subseteq [n]$ be the set of agents which will be successful if they are incentivized. In a greedy algorithm, if the greedy choice is some agent $i \in [n] \setminus T$, since the agent is unsuccessful, the payment is not deducted from the budget. Thus, as a greedy algorithm is not impacted by unsuccessful agents, an adaptive greedy algorithm over $T$ is identical to the sequence of agents selected by running a greedy algorithm directly over $T$.

    Since $\frac{c_i}{p_i}=1$, the payment for each agent is identical. Therefore, conditioned on $T$, the full knowledge problem is simply monotone submodular maximization under a cardinality constraint. From the standard greedy guarantee \cite{NWF78}, the value obtained by greedy is at least $(1-\frac{1}{e})$ times the full knowledge optimum for every realization $T$. Taking expectation over $T$ proves the claim.
    \end{proof}

We next consider agents which require arbitrary payments $t_i$ in order to incentivize effort.
Recall that $OPT_{\mathrm{FK}}$ is the expected value of the optimal algorithm given the full knowledge benchmark. For a realization $T$ of the successful agents, let $OPT(T)$ denote the value of an optimal feasible subset of $T$, let $Singleton(T)$ denote the value of the highest-value feasible singleton in $T$, or $f(\emptyset)$ if none exists, and let $Greedy(T)$ be the value of the standard marginal-density greedy algorithm on $T$.

A known guarantee for monotone submodular maximization under a knapsack constraint \cite{KMN99} restricted to $T$ is that: $$\max\{Singleton(T), Greedy(T)\} \geq \frac{1}{2}(1-\frac{1}{e})OPT(T).$$
In particular:
$$Singleton(T) + Greedy(T) \geq \frac{1}{2} (1-\frac{1}{e})OPT(T).$$

We can therefore make use of the following algorithm:

\begin{algorithm}[h]
\caption{$\frac{1}{4}(1-\frac{1}{e})$-approximation for monotone submodular valuation}
\label{alg:direct_stoch_greedy}
\begin{algorithmic}[1]
\STATE{With probability $\frac{1}{2}$ choose the Singleton strategy, otherwise choose the Greedy strategy.}
\IF{Singleton strategy is chosen}
\WHILE{there exists an untried feasible agent $i$}
\STATE Incentivize the feasible agent $i$ maximizing $f(\{i\})$.
\IF{$i$ is successful}
\STATE Halt.
\ENDIF
\ENDWHILE
\ELSE
\STATE Let $S=\emptyset$.
\WHILE{there exists an untried feasible agent $i$}
\STATE Let $i$ maximize $\frac{f(S\cup\{i\})-f(S)}{t_i}$ among feasible agents.
\STATE Incentivize $i$.
\IF{$i$ is successful}
\STATE $S\leftarrow S\cup\{i\}$.
\ENDIF
\ENDWHILE
\ENDIF
\end{algorithmic}
\end{algorithm}
\begin{theorem}
\label{thm:greedy_submod}
Algorithm \ref{alg:direct_stoch_greedy} is a $\frac{1}{4}(1-\frac{1}{e})$-approximation for $OPT_{\mathrm{FK}}$.
\end{theorem}
\begin{proof}
    Fix a realization $T$. In the singleton strategy, unsuccessful agents consume no budget, so the algorithm considers agents $i$ in decreasing order of $f(\{i\})$ until the highest value successful feasible agent is reached. Its value is therefore exactly $Singleton(T)$.
 
    Similarly, in the greedy strategy, attempting an unsuccessful agent does not consume any of the budget, and does not change the set $S$. Therefore, any unsuccessful agents can simply be ignored. The sequence of successful agents selected by the adaptive algorithm is therefore exactly the sequence selected by running the marginal-density greedy algorithm directly on $T$, and its value is $Greedy(T)$.

    By taking the expectation over $T$ and the initial random choice between the singleton strategy and the greedy strategy, we get:
        $ALG =\frac{1}{2}\EX_T[Singleton(T)] + \frac{1}{2}\EX_T[Greedy(T)] \geq \frac{1}{4}(1-\frac{1}{e})\EX_T[OPT(T)] = \frac{1}{4}(1 - \frac{1}{e})OPT_{\mathrm{FK}}$.
\end{proof}

\subsection{General Binary-Outcome Agents}
We can extend our results for monotone submodular functions beyond binary actions, and also remove the risk-free assumption. However, this requires a different approach, and results in a smaller constant in the approximation ratio. Additionally, the results are no longer an approximation for $OPT_{\mathrm{FK}}$. We do this using the stochastic probing framework of \cite{FSZ21}. In stochastic probing, there is a ground set of elements, where each element is independently active with a possibly distinct known probability. A feasible solution probes elements one at a time, where probing reveals whether an element is active. The set of probed elements $Q$ must satisfy an \emph{outer} feasibility constraint, while the subset $S \subseteq Q$ of probed elements that turn out to be active must satisfy an \emph{inner} feasibility constraint.

\begin{theorem}
\label{thm:stochastic_probing}
For binary-outcome agents with an arbitrary number of actions and a monotone submodular valuation, there is a polynomial-time constant-factor approximation to $OPT$.
\end{theorem}
\begin{proof}
We first justify restricting the reduction to contracts that pay only for success. Any binary-outcome contract $(t_1,t_2)$ can be replaced by
$(t_1-\min\{t_1,t_2\},\,t_2-\min\{t_1,t_2\})$
without changing any incentive constraint and while weakly decreasing every payment. Thus, it is sufficient to consider contracts that only pay for a single outcome. Moreover, it is clearly sufficient to only consider contracts which pay only for success, as otherwise from incentive compatibility a zero-cost action has at least as high a success probability, and can be incentivized using the zero-contract.

For every agent $i$ and action $a\in[\ell]$ that can be induced by a success payment of at most $B$, let $\pi_{i,a}$ be the minimum such payment, and associate a stochastic probing element $(i,a)$ with this agent-action pair. Note that $\pi_{i,a}$ can be computed in polynomial time. Our outer feasibility constraint is that for every agent $i$, $|\{(i,a) \in Q \mid a \in [\ell]\}| \leq 1$, which guarantees that we incentivize at most one action for each agent. Our inner feasibility constraint is that $\sum_{(i,a) \in S} \pi_{i,a} \leq B$, which guarantees that the solution is ex-post budget feasible. Since the action induced by the zero contract corresponds to a probing element with zero payment, any probing solution can be augmented by probing this element for every otherwise unprobed agent after all other probes have been completed, without violating feasibility and while only increasing the value.

The element $(i,a)$ is independently active with probability $p^i_{a,2}$, which corresponds to agent $i$ succeeding when incentivized to choose action $a$. The value of an active set $S$ is then
\[
f\big(\{i\mid (i,a)\in S\text{ for some }a\in[\ell]\}\big),
\]
which is monotone submodular and matches our objective. Thus, the resulting stochastic probing instance is equivalent to our contracting problem. The outer constraint is a partition matroid, while the inner constraint is a knapsack constraint. We can now use the following theorem from \cite{FSZ21}:

\begin{theorem*}[{\cite[Theorem~1.19]{FSZ21}}]
For monotone submodular stochastic probing, a constant-factor approximation can be obtained for a broad class of inner and outer feasibility constraints. 
\end{theorem*}

In our reduction, the outer feasibility constraint is a partition matroid and the inner feasibility constraint is a single knapsack constraint. By \cite[Theorem~1.8]{FSZ21}, matroid and knapsack constraints are among the constraint classes satisfying the requirements of \cite[Theorem~1.19]{FSZ21}. Therefore, the theorem yields the claimed polynomial-time constant-factor approximation.
\end{proof}

\section*{AI Disclosure}
Some of the proofs and some counterexamples in this paper were developed with the assistance of OpenAI's ChatGPT and Anthropic's Claude. 
All the results were independently
 verified and rewritten by the authors.

\section*{Acknowledgments}
This work received funding from the European Research Council (ERC) under the European Union's Horizon 2020 research and innovation program (grant agreement No.~101077862) and the Horizon Europe program for research and innovation (grant agreement No.~101170373), by the Israel Science Foundation (grant No.~3331/24), by the Israel Science Foundation Breakthrough Program (grant No.~2600/24), by the NSF-BSF (grant No.~2020788 and grant No.~2021680), by a Google Research Scholar Award, and by an Amazon Research Award. Vasilis Gkatzelis was supported by NSF CAREER award CCF-2047907 and NSF grant CCF-2210502.

\bibliographystyle{splncs04}
\bibliography{bibliography}
\end{document}